\documentclass[twocolumn]{aastex631}
\usepackage{graphicx}
\usepackage{subfigure}
\usepackage{xcolor,colortbl}

\newcounter{magicrownumbers}
\newcommand\rownumber{\stepcounter{magicrownumbers}\arabic{magicrownumbers}}
\newcolumntype{H}{>{\setbox0=\hbox\bgroup}c<{\egroup}@{}}

\shorttitle{VLA FRAMEx~II: Spectral Results}
\shortauthors{Sargent et al.}

\defcitealias{2021ApJ...906...88F}{FRAMEx~I}
\defcitealias{2022ApJ...927...18F}{FRAMEx~II}
\defcitealias{2022ApJ...936...76S}{FRAMEx~III}
\defcitealias{2023ApJ...958...61F}{FRAMEx~IV}
\defcitealias{2024ApJ...961..109S}{FRAMEx~V}
\defcitealias{2024ApJ...961..230S}{Paper~I}

\graphicspath{{./}{figures/}}

\begin{document}

\title{VLA FRAMEx. II. Radio Spectra of Nearby Active Galactic Nuclei at Subarcsecond Resolution}

\email{andrew.j.sargent2.civ@us.navy.mil}
\AuthorCollaborationLimit=8

\author[0000-0002-8049-0905]{Andrew J.\ Sargent}
\affiliation{United States Naval Observatory, 3450 Massachusetts Ave., NW, Washington, DC 20392, USA}
\affiliation{Department of Physics, The George Washington University, 725 21st St. NW, Washington, DC 20052, USA}

\author[0000-0001-9149-6707]{Alexander J.\ van der Horst}
\affiliation{Department of Physics, The George Washington University, 725 21st St. NW, Washington, DC 20052, USA}

\author[0000-0002-4146-1618]{Megan C. Johnson}
\affiliation{National Science Foundation, 2415 Eisenhower Ave., Alexandria, VA 22314, USA}

\author[0000-0002-3365-8875]{Travis C.\ Fischer}
\affiliation{AURA for ESA, Space Telescope Science Institute, 3700 San Martin Drive, Baltimore, MD 21218, USA}

\author[0000-0002-4902-8077]{Nathan J.\ Secrest}
\affiliation{United States Naval Observatory, 3450 Massachusetts Ave., NW, Washington, DC 20392, USA}

\author[0000-0002-8736-2463]{Phil J.\ Cigan}
\affiliation{United States Naval Observatory, 3450 Massachusetts Ave., NW, Washington, DC 20392, USA}

\author[0000-0003-4727-2209]{Onic I. Shuvo}
\affiliation{Department of Physics, University of Maryland Baltimore County, 1000 Hilltop Circle, Baltimore, MD 21250, USA}

\author[0000-0001-5785-7038]{Krista L.\ Smith}
\affiliation{Department of Physics and Astronomy, Texas A\&M University, 578 University Drive, College Station, TX 77843, USA}

\begin{abstract}
We present $4-12$ GHz in-band spectral energy distributions with accompanying 6 GHz and 10 GHz imaging results for a volume-complete sample ($<40$ Mpc) of hard X-ray selected active galactic nuclei (AGNs) observed with the Karl G. Jansky Very Large Array (VLA) in its A-array configuration.  Despite expectations, only 12 out of 25 of these targets have been detected by the Very Long Baseline Array (VLBA) at milliarcsecond resolution in our previous studies, and we aim to understand the nature of why the circumnuclear radio emission resolves away at the subparsec spatial scales. We find that the sources not detected by the VLBA are also the faintest sources observed with the VLA. We explore the spectral structure derived from the nuclear emission and measure a mean spectral index of $\langle\alpha\rangle=-0.69$ with a scatter of $\sigma_\alpha=0.18$ for the sources not detected by the VLBA, indicative of optically thin synchrotron emission. The 12 sources detected by the VLBA primarily have flat ($-0.5\leq\alpha\leq0.0$) or inverted ($\alpha>0$) spectral indices. Nine of the sources have statistically significant curvature, with only one that was not detected by the VLBA. In NGC~3079, we model an approximately flat spectrum for the excess emission observed by the VLA that is likely produced entirely beyond parsec spatial scales.
\end{abstract}

\section{Introduction} \label{sec:intro}

\begin{deluxetable}{rrRR}[htp!]
\tabletypesize{\footnotesize}
\tablecaption{VLA FRAMEx Sample}
\tablehead{
&
\colhead{Target} &
\colhead{R.A. (ICRS)} &
\colhead{Decl. (ICRS)}
}
\startdata
 \setcounter{magicrownumbers}{0}
 \rownumber & NGC~1052 & $02^{\rm h}41^{\rm m}04.790^{\rm s}$ & $-08\degr15\arcmin20.70\arcsec$  \\ 
 \rownumber & NGC~1068 & $02^{\rm h}42^{\rm m}40.770^{\rm s}$ & $-00\degr00\arcmin47.80\arcsec$  \\ 
 \rownumber & NGC~1320 & $03^{\rm h}24^{\rm m}48.690^{\rm s}$ & $-03\degr02\arcmin32.10\arcsec$  \\ 
 \rownumber & NGC~2110 & $05^{\rm h}52^{\rm m}11.370^{\rm s}$ & $-07\degr27\arcmin22.40\arcsec$  \\ 
 \rownumber & NGC~2782 & $09^{\rm h}14^{\rm m}05.110^{\rm s}$ & $+40\degr06\arcmin49.60\arcsec$  \\ 
 \rownumber &  IC~2461 & $09^{\rm h}19^{\rm m}58.030^{\rm s}$ & $+37\degr11\arcmin27.70\arcsec$  \\ 
 \rownumber & NGC~2992 & $09^{\rm h}45^{\rm m}42.040^{\rm s}$ & $-14\degr19\arcmin34.80\arcsec$  \\ 
 \rownumber & NGC~3081 & $09^{\rm h}59^{\rm m}29.540^{\rm s}$ & $-22\degr49\arcmin34.70\arcsec$  \\ 
 \rownumber & NGC~3089 & $09^{\rm h}59^{\rm m}36.680^{\rm s}$ & $-28\degr19\arcmin52.70\arcsec$  \\ 
 \rownumber & NGC~3079 & $10^{\rm h}01^{\rm m}57.800^{\rm s}$ & $+55\degr40\arcmin47.20\arcsec$  \\ 
 \rownumber & NGC~3227 & $10^{\rm h}23^{\rm m}30.570^{\rm s}$ & $+19\degr51\arcmin54.20\arcsec$  \\ 
 \rownumber & NGC~3786 & $11^{\rm h}39^{\rm m}42.510^{\rm s}$ & $+31\degr54\arcmin33.90\arcsec$  \\ 
 \rownumber & NGC~4151 & $12^{\rm h}10^{\rm m}32.570^{\rm s}$ & $+39\degr24\arcmin21.00\arcsec$  \\ 
 \rownumber & NGC~4180 & $12^{\rm h}13^{\rm m}03.050^{\rm s}$ & $+07\degr02\arcmin19.60\arcsec$  \\ 
 \rownumber & NGC~4235 & $12^{\rm h}17^{\rm m}09.880^{\rm s}$ & $+07\degr11\arcmin29.60\arcsec$  \\ 
 \rownumber & NGC~4388 & $12^{\rm h}25^{\rm m}46.810^{\rm s}$ & $+12\degr39\arcmin43.40\arcsec$  \\ 
 \rownumber & NGC~4593 & $12^{\rm h}39^{\rm m}39.440^{\rm s}$ & $-05\degr20\arcmin39.00\arcsec$  \\ 
 \rownumber & NGC~5290 & $13^{\rm h}45^{\rm m}19.160^{\rm s}$ & $+41\degr42\arcmin44.40\arcsec$  \\ 
 \rownumber & NGC~5506 & $14^{\rm h}13^{\rm m}14.900^{\rm s}$ & $-03\degr12\arcmin27.20\arcsec$  \\ 
 \rownumber & NGC~5899 & $15^{\rm h}15^{\rm m}03.250^{\rm s}$ & $+42\degr02\arcmin59.40\arcsec$  \\ 
 \rownumber & NGC~6814 & $19^{\rm h}42^{\rm m}40.580^{\rm s}$ & $-10\degr19\arcmin25.10\arcsec$  \\ 
 \rownumber & NGC~7314 & $22^{\rm h}35^{\rm m}46.190^{\rm s}$ & $-26\degr03\arcmin01.50\arcsec$  \\ 
 \rownumber & NGC~7378 & $22^{\rm h}47^{\rm m}47.690^{\rm s}$ & $-11\degr48\arcmin59.80\arcsec$  \\ 
 \rownumber & NGC~7465 & $23^{\rm h}02^{\rm m}00.950^{\rm s}$ & $+15\degr57\arcmin53.50\arcsec$  \\ 
 \rownumber & NGC~7479 & $23^{\rm h}04^{\rm m}56.660^{\rm s}$ & $+12\degr19\arcmin22.30\arcsec$     
\enddata
\tablecomments{The VLA FRAMEx sample and their ICRS coordinates used for pointing in observations.}
\label{tab:sample}
\end{deluxetable}

Dynamical evidence has shown that supermassive black holes (SMBHs, with masses $>10^6M_{\odot}$) with deep gravitational potentials reside at the centers of massive galaxies \citep{1995ARA&A..33..581K}. When matter falls onto SMBHs in the form of an accretion disk, radiation is emitted across the electromagnetic spectrum and the host galaxy nuclei become active \citep{1969Natur.223..690L}. The accretion process regulates the surrounding interstellar medium (ISM) in these active galactic nuclei (AGNs) through physical and radiative mechanisms strongly correlated with radio luminosity at GHz frequencies \citep{1978A&A....64..433D,2012ARA&A..50..455F,2019NatAs...3..387P}. Over time, nearby gas and dust is cleared, depriving the SMBH of necessary fuel and terminating any further AGN activity.

Measurement of the radio continuum is essential when characterizing nuclear properties of AGNs, most importantly because it is not susceptible to the effects of dust obscuration that are prominent at higher frequencies \citep{2001ApJS..133...77H}. An important diagnostic is the shape of the spectral energy distribution (SED), as a variety of physical mechanisms produce the radio spectral index $\alpha$ (defined as $S\propto\nu^{\alpha}$), which is typically classified as steep ($\alpha<-0.5$), flat ($-0.5 \leq \alpha \leq 0$), or inverted ($\alpha>0$). Flat or inverted spectra may be due to attenuation from optically thick sources observed near the nucleus. For example, distributions of electrons with overlapping synchrotron spectra at radii near the base of a highly collimated relativistic jet may produce flat to inverted spectra observed along the line of sight \citep[$-0.5 < \alpha < 2.5$,][]{1979ApJ...232...34B}. Electrons accelerated in a highly magnetized corona can also contribute to forming flat spectra through non-uniform superpositions of many synchrotron sources which are self-absorbed \citep{1976A&A....52..439D,2008MNRAS.390..847L}. Steep spectrum sources on the other hand are observed in optically thin emission regions. This emission may originate from AGN winds shocking the ISM \citep{2014MNRAS.442..784Z}, old electron populations ($\gtrsim 10^7~{\rm yr}$) within remnants of supernovae events \citep{1992ARA&A..30..575C}, or relativistic electrons accelerating along magnetized plasma in a collimated jet \citep{1984RvMP...56..255B}. The radio spectrum is also often accompanied by free-free radiation, with optically thick nuclear sources potentially producing an inverted spectrum \citep{2021MNRAS.508..680B,2019MNRAS.482.5513L,1995ApJ...448..589L}, or the optically thin flat-spectrum \ion{H}{2} regions associated with star formation \citep[][]{1992ARA&A..30..575C}.

Measuring and interpreting the radio spectral index can be challenging. Sampling from heterogeneous surveys can complicate the determination of radio spectra, where flux density measurements may suffer from unmatched resolutions, differing $(u,~v)$-coverage, and non-simultaneity, all of which can potentially skew the spectral index measurement \citep[see for instance Figure 11 in][]{2019ApJ...875...80G}.
Further restrictions apply when sensitivity preferences need to be reduced due to telescope observing time, data volume, or computational limitations. Lastly and depending on the science goals, sample selection biases, including wavelength-dependent selection effects and incomplete galaxy samples, can complicate the statistical interpretation of spectral indices.

Nearby AGNs have frequently been observed with the Very Large Array (VLA), a 27-dish radio interferometer constructed in the 1970s, which has been used to determine their spectral properties \cite[e.g.][]{1996ApJ...473..130R,2001ApJS..133...77H}. But a complicating issue common in the decades-old archival observations is sparse spectral coverage, often with only a few narrow-band measurements that span large frequency ranges due to the receiver limitations at the time. However, in 2012 the VLA was upgraded to the Karl G. Jansky VLA (hereafter VLA), mitigating several of the limitations. The upgrade expanded sensitivity by at least an order of magnitude from several perspectives including the continuum sensitivity ($\times10$), bandwidth ($\times80$), and fine frequency resolution ($\times3180)$. Additionally, the frequency coverage is now complete from $1-50~{\rm GHz}$, and allows for simultaneous sampling across 4 GHz with all polarization correlations. Therefore, the VLA offers unique advantages for constructing AGN SEDs across large frequency ranges while maintaining consistent spatial scales due to the flexibility provided by four primary antenna configurations. This was recently exemplified in \cite{2022MNRAS.515..473P}, a multifrequency study using multiple VLA configurations to form matched resolution SEDs across $5-45~{\rm GHz}$ from a sample of hard X-ray selected AGNs.

The Fundamental Reference AGN Monitoring Experiment project \citep[FRAMEx;][]{2020jsrs.conf..165D} is an observing program studying the nuclear phenomena of a hard X-ray selected volume-complete sample of 25 AGNs in the time domain. The FRAMEx sample was chosen in part to study the implied relation between the radio and X-ray luminosities observed in the fundamental plane of black hole activity (FP), that, when combined with the black hole mass, defines a paradigm unifying the ostensible core emission processes for black holes of all masses \citep{2003MNRAS.345.1057M,2004A&A...414..895F}. We obtained high-resolution ($\sim$mas) Very Long Baseline Array (VLBA) snapshot observations together with \textit{Neil Gehrels Swift Observatory (Swift)} X-ray Telescope (XRT) observations in \citet[][hereafter \citetalias{2021ApJ...906...88F}]{2021ApJ...906...88F} and used the FP to derive an expected radio luminosity with a $5\sigma$ detection threshold. Despite a 100\% detection rate of the previously available archival VLA observations of the FRAMEx targets from the NRAO VLA Archive Survey\footnote{\url{http://archive.nrao.edu/nvas}} \citepalias[depicted in Figure 3 in][]{2021ApJ...906...88F}, detected only 9 out of 25 targets with the VLBA, an unexpected result that implies an inconsistency with the expectations of the FP relation. A reobservation campaign subsequently occurred and focused on 9 of the 16 objects that were not originally detected with the VLBA, but with four times the integration time in order to double the sensitivity as compared to the original observation \citep[][hereafter \citetalias{2022ApJ...936...76S}]{2022ApJ...936...76S}. Still, only 3 of the 9 targets were detected with these deeper observations, bringing the total number of FRAMEx detections to 12 out of 25. Two additional VLBA campaigns followed the snapshot observations in \citetalias{2021ApJ...906...88F} which focused on the VLBA-detected targets. One was a monitoring campaign studying the parsec-scale variability (i.e. the original intent of FRAMEx) in the targets NGC~2992 \citep[][named \citetalias{2022ApJ...927...18F}]{2022ApJ...927...18F} and NGC~3079 \citep[][hereafter \citetalias{2023ApJ...958...61F}]{2023ApJ...958...61F}. \citet[][hereafter \citetalias{2024ApJ...961..109S}]{2024ApJ...961..109S} conducted multifrequency observations of all of the VLBA-detected objects at 1.6, 4.4, 8.6, and 22 GHz to more robustly determine the true central origin of radio emission.

The discrepancy with the FP suggests that much of the radio emission observed to be coincident with the nucleus in the archival VLA data is contaminated with extranuclear radio emission that is produced beyond the resolving capabilities of the VLBA. For reference, a 6 GHz VLA observation in its most extended A-array configuration has a maximum baseline length of 36 km which sets a resolution of $\sim0.3''$. Likewise, the smallest baseline length of 0.68 km of the VLA can resolve emission out to a largest angular scale of $\sim9''$. Thus, at a luminosity distance of 40 Mpc the VLA can resolve emission that spans projected physical sizes of $\sim58-1745$ pc. Compare this to the corresponding resolving capabilities of the VLBA set by its shortest (Pie~Town--Los~Alamos, 236 km) and longest baselines (Mauna~Kea--St.~Croix, separated by 8611 km\footnote{\url{https://science.nrao.edu/facilities/vlba/docs/manuals/oss/ang-res}}), which probe angular scales of $\sim0.001''-0.004''$, corresponding to regions extending $\sim0.38-10$ pc in physical sizes. The VLBA is therefore most sensitive to compact point sources and does not have the ability to resolve extended emission very well beyond the largest angular scale constrained by the shortest baseline. The VLBA sensitivity is further exacerbated by its sparse 10-antenna array, which provides for 45 baselines, and our one-hour snapshot observations in \citetalias{2021ApJ...906...88F} probed to an observation sensitivity depth of $\sim20~{\rm \mu Jy~bm^{-1}}$ with the full array. On the other hand, a theoretical one-hour observation with all 27 antennas (351 baselines) of the VLA achieves a sensitivity depth of $\sim7~{\rm \mu Jy~bm^{-1}}$.

Therefore, this paper investigates the Stokes $I$ radio continuum emission as a follow-up survey to \citetalias{2021ApJ...906...88F} with new VLA observations. The observations consist of a uniform data set of the FRAMEx sample observed with the VLA in its most extended A-array configuration, resulting in the highest spatial resolution achievable with the VLA. It is also follow-up to \citet[][hereafter \citetalias{2024ApJ...961..230S}]{2024ApJ...961..230S}, where we focused on the methodology of calibration and imaging for these data, and conducted an in-depth analysis of one of the 25 sources in our sample, NGC~4388. Here we conduct an analysis of the $4-12~{\rm GHz}$ spectral properties of the AGNs with a $128~{\rm MHz}$ spectral resolution. In Section \ref{sec:methodology}, we discuss the sample, calibration, and imaging procedures used for these new VLA observations. In Sections \ref{sec:results} and \ref{sec:discussion} we present and discuss our results, and in Section \ref{sec:conclusion} we summarize our findings.

Throughout this paper we define the spectral index $\alpha$ as $S\propto\nu^{\alpha}$ and spectral curvature $\beta$ as $S\propto\nu^{(\alpha+\beta\log\nu)}$, with $S$ the flux density and $\nu$ the frequency.

\begin{deluxetable*}{rrcclc|lcc|r}
\tabletypesize{\footnotesize}
\tablecaption{VLA FRAMEx Observations}
\tablehead{
&
\colhead{Target} &
\colhead{Band} &
\colhead{Obs. Date} &
\colhead{Conf.} &
\colhead{SB ID} &
\colhead{Primary} &
\colhead{Interpolated} &
\colhead{Model} &
\colhead{Phase}
\\[-0.1cm]
\colhead{} &
\colhead{} &
\colhead{} &
\colhead{} &
\colhead{} &
\colhead{} &
\colhead{Cal.} &
\colhead{Flux} &
\colhead{Diff.} &
\colhead{Cal.}
\\[-0.1cm]
\colhead{} &
\colhead{(1)} &
\colhead{(2)} &
\colhead{(3)} &
\colhead{(4)} &
\colhead{(5)} &
\colhead{(6)} &
\colhead{(7)} &
\colhead{(8)} &
\colhead{(9)}
}
\startdata
 \setcounter{magicrownumbers}{0}
 \rownumber &          NGC~1052     & C & 2021-01-24 & A                 & 38952694 & 3C~138 & 3.517 Jy & 3.7\%   & J0239$-$0234 \\
            & & X & 2021-01-16 & A                 & 38953167 & 3C~138 & 2.484 Jy & 6.5\%   & J0239$-$0234 \\
 \rownumber &          NGC~1068     & C & 2021-01-26 & A                 & 39008458 & 3C~138 & 3.518 Jy & 3.7\%   & J0239$-$0234 \\
            & & X & 2021-01-20 & A                 & 39008718 & 3C~138 & 2.485 Jy & 6.6\%   & J0239$-$0234 \\
 \rownumber &          NGC~1320     & C & 2021-01-24 & A                 & 39011072 & 3C~138 & 3.517 Jy & 3.7\%   & J0339$-$0146 \\
            & & X & 2021-01-17 & A                 & 39175511 & 3C~138 & 2.484 Jy & 6.6\%   & J0339$-$0146 \\
 \rownumber &          NGC~2110     & C & 2021-02-04 & A                 & 39176337 & 3C~138 & 3.518 Jy & 3.7\% & J0541$-$0541 \\
            & & X & 2021-01-27 & A                 & 39176455 & 3C~138 & 2.488 Jy & 6.7\% & J0541$-$0541 \\
 \rownumber &          NGC~2782     & C & 2020-12-31 & A                 & 38677768 & 3C~138 & 3.515 Jy & 3.6\% & J0927$+$3902 \\
            & & X & 2020-12-14 & A                 & 38952497 & 3C~286 & \nodata  & \nodata & J0927$+$3902 \\
\rownumber &           IC~2461     & C & 2021-02-03 & A                 & 39176799 & 3C~286 & \nodata  & \nodata & J0927$+$3902 \\
            & & X & 2021-01-18 & A                 & 39176643 & 3C~286 & \nodata  & \nodata & J0927$+$3902 \\
 \rownumber &          NGC~2992     & C & 2021-02-19 & A                 & 39236874 & 3C~286 & \nodata  & \nodata & J0943$-$0819 \\
            & & X & 2021-01-29 & A                 & 39237182 & 3C~286 & \nodata  & \nodata & J0943$-$0819 \\
 \rownumber &          NGC~3081     & C & 2021-01-06 & A                 & 38947401 & 3C~138 & 3.516 Jy & 3.6\%   & J0927$-$2034 \\
            & & X & 2021-01-03 & A                 & 38954868 & 3C~138 & 2.479 Jy & 6.3\%   & J0927$-$2034 \\
 \rownumber &          NGC~3089     & C & 2021-01-12 & A                 & 38947689 & 3C~138 & 3.516 Jy & 3.6\%   & J1037$-$2934 \\
            & & X & 2021-01-04 & A                 & 39007522 & 3C~138 & 2.479 Jy & 6.4\%   & J1037$-$2934 \\
 \rownumber &          NGC~3079     & C & 2021-02-25 & A                 & 39342314 & 3C~138 & 3.520 Jy & 3.7\%   & J1035$+$5628 \\
            & & X & 2021-02-24 & A                 & 39342185 & 3C~138 & 2.498 Jy & 7.2\%   & J1035$+$5628 \\
 \rownumber &          NGC~3227     & C & 2021-03-02 & A$\rightarrow$D   & 39343798 & 3C~138 & 3.521 Jy & 3.8\%   & J1016$+$2037 \\
            & & X & 2021-03-01 & A                 & 39343665 & 3C~138 & 2.500 Jy & 7.2\%   & J1016$+$2037 \\
 \rownumber &          NGC~3786     & X & 2021-03-03 & A$\rightarrow$D   & 39380922 & 3C~286 & \nodata  & \nodata & J1147$+$3501 \\
 \rownumber &          NGC~4151     & C & 2021-02-18 & A                 & 39236054 & 3C~286 & \nodata  & \nodata & J1209$+$4119 \\
            & & X & 2021-01-22 & A                 & 39236360 & 3C~286 & \nodata  & \nodata & J1209$+$4119 \\
 \rownumber &          NGC~4180     & C & 2020-12-31 & A                 & 39007910 & 3C~286 & \nodata  & \nodata & J1224$+$0330 \\
            & & X & 2020-12-10 & A                 & 39008114 & 3C~286 & \nodata  & \nodata & J1224$+$0330 \\
 \rownumber &          NGC~4235     & C & 2021-03-03 & A$\rightarrow$D   & 39328466 & 3C~286 & \nodata  & \nodata & J1239$+$0730 \\
            & & X & 2021-03-01 & A                 & 39336923 & 3C~286 & \nodata  & \nodata & J1239$+$0730 \\
 \rownumber &          NGC~4388     & C & 2021-03-03 & A$\rightarrow$D   & 39381834 & 3C~286 & \nodata  & \nodata & J1254$+$1141 \\
            & & X & 2021-03-02 & A$\rightarrow$D   & 39382198 & 3C~286 & \nodata  & \nodata & J1254$+$1141 \\
 \rownumber &          NGC~4593     & C & 2021-01-01 & A                 & 38953664 & 3C~286 & \nodata  & \nodata & J1246$-$0730 \\
            & & X & 2020-12-18 & A                 & 38954482 & 3C~286 & \nodata  & \nodata & J1246$-$0730 \\
 \rownumber &          NGC~5290     & C & 2021-03-01 & A                 & 39375363 & 3C~286 & \nodata  & \nodata & J1327$+$4326 \\
            & & X & 2021-02-28 & A                 & 39367450 & 3C~286 & \nodata  & \nodata & J1327$+$4326 \\
 \rownumber &          NGC~5506     & C & 2021-03-02 & A$\rightarrow$D   & 39365069 & 3C~286 & \nodata  & \nodata & J1354$-$0206 \\
            & & X & 2021-03-01 & A                 & 39364671 & 3C~286 & \nodata  & \nodata & J1354$-$0206 \\
 \rownumber &          NGC~5899     & C & 2021-03-02 & A$\rightarrow$D   & 39375979 & 3C~286 & \nodata  & \nodata & J1506$+$4239 \\
            & & X & 2021-03-01 & A                 & 39375853 & 3C~286 & \nodata  & \nodata & J1506$+$4239 \\
 \rownumber &          NGC~6814     & C & 2021-03-04 & A$\rightarrow$D   & 39380472 & 3C~286 & \nodata  & \nodata & J1939$-$1002 \\
            & & X & 2021-03-04 & A$\rightarrow$D   & 39376467 & 3C~286 & \nodata  & \nodata & J1939$-$1002 \\
 \rownumber &          NGC~7314     & C & 2021-02-27 & A                 & 39338173 & 3C~48  & 4.495 Jy & 1.8\%   & J2258$-$2758 \\
            & & X & 2021-02-28 & A                 & 39339300 & 3C~48  & 2.697 Jy & 0.6\%   & J2258$-$2758 \\
 \rownumber &          NGC~7378     & C & 2021-03-02 & A$\rightarrow$D   & 39341061 & 3C~48  & 4.491 Jy & 1.7\%   & J2246$-$1206 \\
            & & X & 2021-02-28 & A                 & 39341696 & 3C~48  & 2.697 Jy & 0.6\%   & J2246$-$1206 \\
 \rownumber &          NGC~7465     & C & 2021-03-02 & A$\rightarrow$D   & 39337668 & 3C~147 & \nodata  & \nodata & J2253$+$1608 \\
            & & X & 2021-02-28 & A                 & 39337786 & 3C~147 & \nodata  & \nodata & J2253$+$1608 \\
 \rownumber &          NGC~7479     & C & 2021-02-28 & A                 & 39340326 & 3C~48  & 4.494 Jy & 1.7\%   & J2253$+$1608 \\
            & & X & 2021-02-26 & A                 & 39340840 & 3C~147 & \nodata  & \nodata & J2253$+$1608
\enddata
\tablecomments{\textbf{Column 1.} target name. \textbf{Column 2.} frequency band observed. \textbf{Column 3.} date of observation. \textbf{Column 4.} VLA configuration during observation. \textbf{Column 5.} NRAO scheduling block ID. \textbf{Column 6.} primary calibrator. \textbf{Column 7.} interpolated flux density for flaring calibrator at date of observation that primary calibrator was scaled to. \textbf{Column 8.} percent difference of flux density from \cite{2017ApJS..230....7P} as determined by \texttt{getcalmodvla}. \textbf{Column 9.} secondary phase reference calibrator. }
\label{tab:observations}
\end{deluxetable*}

\section{Methodology} \label{sec:methodology}
\subsection{Sample Selection}
The sample in this paper is the same as in \citetalias{2021ApJ...906...88F}, which were hard X-ray selected AGNs from the 105-month ({\it Swift}) Burst Alert Telescope (BAT) catalog \citep{2018ApJS..235....4O}. The catalog's uniform flux limit of ${\sim}8\times10^{-12}~{\rm erg~s^{-1}~cm^{-2}}$ enables a selection of all AGNs above a defined luminosity threshold out to a corresponding volume-complete luminosity distance. Thus the FRAMEx targets were selected with a hard X-ray luminosity ${>}10^{42}~{\rm erg~s^{-1}}$, limiting the distance of the sample to all AGNs within 40~Mpc. The original FRAMEx sample selection imposed declination limits of $-30\degr < \delta < +60\degr$ for observational reasons, yielding 25 objects which form the basis of the present study. Table \ref{tab:sample} lists the sample with their ICRS positions.

\subsection{VLA Observations and Data Reduction}
We conducted VLA observations of our 25 targets between 2020 December 10 and 2021 March 4 under the project code 20B-241 (P.I. Fischer). Most of the observations were carried out when the VLA was in the A-array configuration, but some when the VLA was transitioning to its D-array configuration after 2021 March 1 (11 out of 50 observations were affected by this transition). Only a few antennas (2--4) were transitioned per day during the array maneuver, starting with the antennas within the inner core of the array. Our observations concluded on 2021 March 4 and none of outer baselines were affected, maintaining our desired resolution but increasing the largest angular scales to $\sim240''$ ($\sim47$ kpc) for the affected data sets. Each target was observed twice, with one observation at C band, from 4 to 8 GHz, and the other at X band, from 8 to 12 GHz. For each scheduling block, observations for absolute flux density scaling and bandpass calibrations were conducted on one of the primary calibrators 3C~48, 3C~138, 3C~147, or 3C~286. A secondary calibrator nearby the target was observed by alternating 2-minute observations for phase referencing with 8-minute observations of the target source. Additionally, calibrator sources were observed for the polarization angle and the polarization crossterms with each scheduling block, but the analysis of polarization properties for each target will be presented in a future publication. Table \ref{tab:observations} lists each target's observation date for each band, the associated scheduling block ID and VLA array configuration, as well as the primary and secondary calibrators used.

We used the Common Astronomy Software Applications \citep[\textsc{casa;}][]{2022PASP..134k4501C} software package, version 6.6, to calibrate all data. The data reduction was largely done in the same way as in \citetalias{2024ApJ...961..230S}, with a caveat on the primary calibrator source used for flux density scaling. We first used a~priori measurements to determine an initial delay calibration per antenna relative to a reference antenna. We set the flux density of the primary calibrator with \texttt{setjy} for observations with 3C~147 and 3C~286 using the Perley-Butler 2017 models \citep{2017ApJS..230....7P}. However, for the duration of our project, 3C~48 and 3C~138 were both in a flaring state. According to private communications with National Radio Astronomy Observatory (NRAO), the sources are brighter than the Perley-Butler 2017 models, but the flaring is not occurring on a rapid timescale. NRAO conducts monthly monitoring observations of the calibrator sources across all frequency bands and they provided a function which interpolates the flux density for the calibrators at a given date. The function downloads a list of flux density components which are then Fourier transformed with \texttt{ft} to model the calibrators across each band. Table \ref{tab:observations} lists the interpolated flux densities at 6 GHz and 10 GHz and their percentage differences from the Perley-Butler 2017 models.\footnote{We also checked for flux density differences in 3C~147 but this source showed a 0\% difference from the Perley-Butler 2017 model for both bands.} Once the model was set, each observation was calibrated for the bandpass and the complex gain. The calibrated data were inspected for radio frequency interference (RFI), which was then subsequently removed before recalibrating without the RFI affected channels. Once the data were calibrated, the target sources were separated from the measurement sets using \texttt{split} and \texttt{statwt} was run to reweight the data based on its variance.

\begin{figure*}[ht!]
    \centering
    \includegraphics[width=\textwidth]{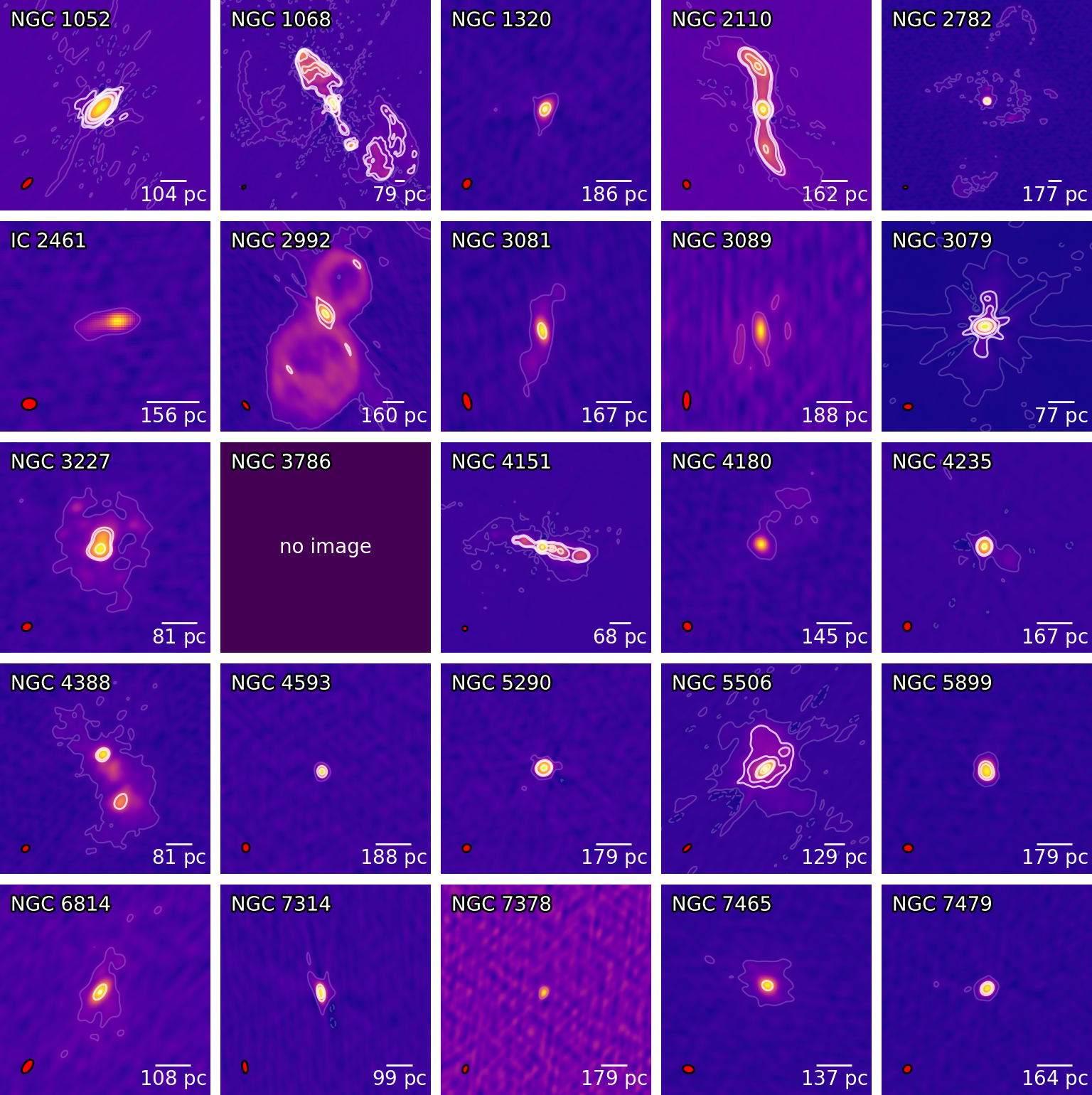}
    \caption{6 GHz MTMFS imaging results for the VLA FRAMEx sample.  The thin contour lines represent $\pm5\times$rms noise for the respective target (see Table \ref{tab:sensitivity}). The thick contour lines represent 0.5, 1, 5, 10, 50, and 100 ${\rm mJy~bm^{-1}}$. The restoring beams are represented by the red ellipse in the bottom left corner. The scale bar represents an angular size of $1\arcsec$ and indicates the corresponding length projection.}
    \label{fig:vlatargets_C}
\end{figure*}

\begin{figure*}[ht!]
    \centering
    \includegraphics[width=\textwidth]{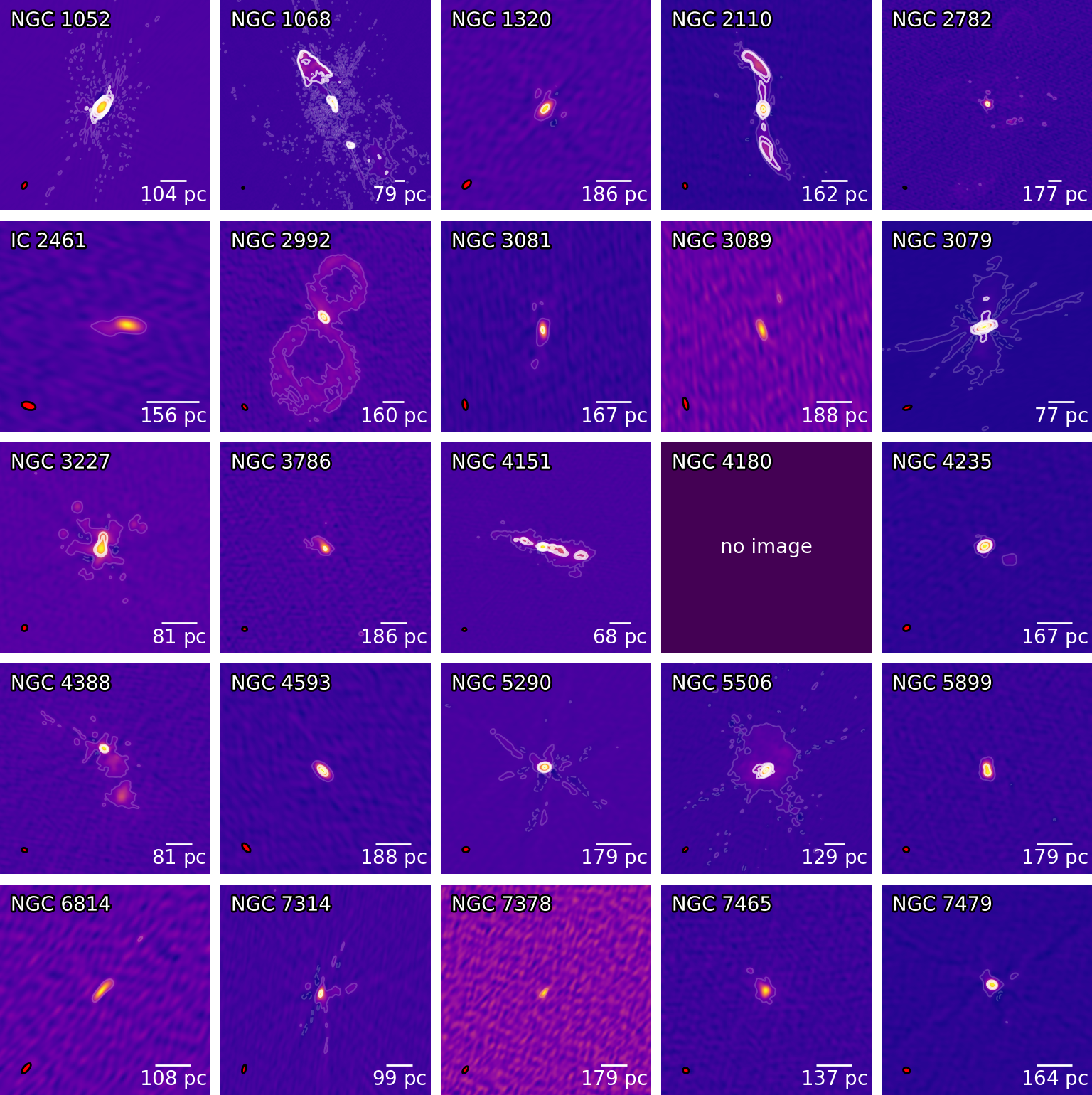}
    \caption{10 GHz MTMFS imaging results for the VLA FRAMEx sample. The thin contour lines represent $\pm5\times$rms noise for the respective target (see Table \ref{tab:sensitivity}). The thick contour lines represent 0.5, 1, 5, 10, 50, and 100 ${\rm mJy~bm^{-1}}$. The restoring beams are represented by the red ellipse in the bottom left corner. The scale bar represents an angular size of $1\arcsec$ and indicates the corresponding length projection.}
    \label{fig:vlatargets_X}
\end{figure*}

\begin{deluxetable*}{rrccCCccc}
\tabletypesize{\footnotesize}
\tablecaption{Sensitivity}
\tablehead{
&
\colhead{Target} &
\colhead{Band} &
\colhead{Exposure} &
\colhead{Restoring Beam} &
\colhead{Position Angle} &
\colhead{$\sigma_{\rm theor.}$} &
\colhead{$\sigma_{\rm rms}$} &
\colhead{Self-Cal.} 
\\[-0.1cm]
&
&
&
\colhead{(s)} &
(\rm{arcsec $\times$ arcsec})&
(\rm{degree}) &
\colhead{($\rm \mu Jy~bm^{-1}$)} &
\colhead{($\rm \mu Jy~bm^{-1}$)} &
\\[-0.1cm]
&
\colhead{(1)} &
\colhead{(2)} &
\colhead{(3)} &
\colhead{(4)} &
\colhead{(5)} &
\colhead{(6)} &
\colhead{(7)} &
\colhead{(8)}
}
\startdata
 \setcounter{magicrownumbers}{0}
 \rownumber &          NGC~1052     & C & 2564    & 0.53   \times 0.25   & -45.1  & 4.01     & 30.73   & p+a \\
            & & X & 2564    & 0.28   \times 0.16   & -34.2  & 3.06     & 21.41   & p+a \\
 \rownumber &          NGC~1068     & C & 2860    & 0.40   \times 0.25   & -45.4  & 3.75     & 12.76   & p+a \\
            & & X & 2860    & 0.18   \times 0.18   & -35.5  & 3.06     & 11.40   & p+a \\
 \rownumber &          NGC~1320     & C & 2560    & 0.31   \times 0.24   & -29.8  & 3.80     & 6.42    & \nodata \\ 
            & & X & 2564    & 0.30   \times 0.17   & -45.5  & 3.20     & 7.92    & \nodata \\ 
 \rownumber &          NGC~2110     & C & 2558    & 0.34   \times 0.27   & 20.5   & 4.04     & 6.80    & p+a \\ 
            & & X & 2374    & 0.23   \times 0.17   & 12.2   & 3.03     & 16.52   & p+a \\ 
 \rownumber &          NGC~2782     & C & 2382    & 0.33   \times 0.24   & -84.0  & 3.62     & 6.25    & \nodata \\
            & & X & 2384    & 0.25   \times 0.16   & 72.1   & 3.16     & 5.54    & \nodata \\
 \rownumber &           IC~2461     & C & 2560    & 0.29   \times 0.23   & -83.4  & 3.56     & 6.53    & \nodata \\ 
            & & X & 2560    & 0.27   \times 0.14   & 74.5   & 2.68     & 5.41    & \nodata \\ 
 \rownumber &          NGC~2992     & C & 2860    & 0.50   \times 0.24   & 36.7   & 3.43     & 5.85    & \nodata \\
            & & X & 2860    & 0.31   \times 0.19   & 39.9   & 2.56     & 8.58    & p \\
 \rownumber &          NGC~3081     & C & 2860    & 0.49   \times 0.23   & 15.6   & 3.63     & 6.83    & \nodata \\
            & & X & 2860    & 0.30   \times 0.14   & 9.6    & 2.91     & 5.48    & \nodata \\
 \rownumber &          NGC~3089     & C & 2322    & 0.54   \times 0.21   & -0.8   & 4.03     & 6.73    & \nodata \\
            & & X & 2800    & 0.36   \times 0.13   & 13.8   & 2.89     & 6.03    & \nodata \\
 \rownumber &          NGC~3079     & C & 2384    & 0.36   \times 0.25   & -84.6  & 3.68     & 7.05    & p+a \\
            & & X & 1906    & 0.33   \times 0.14   & -71.1  & 3.16     & 8.69    & p+a \\
 \rownumber &          NGC~3227     & C & 2620    & 0.30   \times 0.23   & -68.3  & 4.28     & 6.77    & \nodata \\
            & & X & 2624    & 0.18   \times 0.16   & -33.7  & 2.78     & 5.34    & p \\
 \rownumber &          NGC~3786     & X & 2626    & 0.19   \times 0.15   & 88.2   & 2.96     & 4.93    & \nodata \\
 \rownumber &          NGC~4151     & C & 2862    & 0.24   \times 0.23   & 34.3   & 3.42     & 5.01    & p+a \\
            & & X & 2862    & 0.19   \times 0.14   & -84.9  & 2.68     & 5.50    & p+a \\ 
 \rownumber &          NGC~4180     & C & 2860    & 0.28   \times 0.25   & 36.8   & 3.43     & 5.41    & \nodata \\
            & & X & \nodata & \nodata              & \nodata & \nodata & \nodata & \nodata \\
 \rownumber &          NGC~4235     & C & 2860    & 0.27   \times 0.23   & -9.7   & 3.63     & 5.04    & \nodata \\
            & & X & 2860    & 0.21   \times 0.16   & -58.6  & 2.77     & 4.89    & p \\
 \rownumber &          NGC~4388     & C & 2622    & 0.32   \times 0.25   & -53.4  & 3.72     & 5.48    & \nodata \\
            & & X & 2626    & 0.22   \times 0.16   & 71.6   & 2.91     & 6.05    & \nodata \\
 \rownumber &          NGC~4593     & C & 2862    & 0.26   \times 0.21   & 2.2    & 3.86     & 18.66   & \nodata \\
            & & X & 2860    & 0.31   \times 0.15   & 41.7   & 2.79     & 6.11    & \nodata \\
 \rownumber &          NGC~5290     & C & 2624    & 0.25   \times 0.22   & -56.2  & 3.72     & 6.03    & \nodata \\
            & & X & 2622    & 0.19   \times 0.15   & -83.3  & 2.69     & 5.14    & \nodata \\
 \rownumber &          NGC~5506     & C & 2620    & 0.50   \times 0.23   & -47.2  & 4.12     & 7.70    & p \\
            & & X & 2622    & 0.29   \times 0.16   & -44.4  & 3.50     & 8.02    & p \\
 \rownumber &          NGC~5899     & C & 2620    & 0.28   \times 0.23   & 86.5   & 3.87     & 5.51    & \nodata \\
            & & X & 2620    & 0.17   \times 0.15   & 74.2   & 2.80     & 4.67    & \nodata \\
 \rownumber &          NGC~6814     & C & 2624    & 0.45   \times 0.24   & -37.1  & 3.54     & 6.25    & \nodata \\
            & & X & 2622    & 0.35   \times 0.15   & -41.2  & 4.09     & 16.33   & \nodata \\
 \rownumber &          NGC~7314     & C & 3042    & 0.48   \times 0.21   & 7.8    & 3.39     & 5.65    & \nodata \\
            & & X & 3040    & 0.33   \times 0.13   & -14.3  & 2.50     & 5.02    & \nodata \\
 \rownumber &          NGC~7378     & C & 3040    & 0.33   \times 0.21   & -19.5  & 4.07     & 6.43    & \nodata \\
            & & X & 3040    & 0.31   \times 0.15   & -36.1  & 2.73     & 5.20    & \nodata \\
 \rownumber &          NGC~7465     & C & 2860    & 0.32   \times 0.23   & 78.1   & 4.27     & 6.20    & \nodata \\
            & & X & 2860    & 0.17   \times 0.15   & 64.4   & 2.68     & 4.87    & \nodata \\
 \rownumber &          NGC~7479     & C & 2858    & 0.26   \times 0.22   & -44.0  & 3.86     & 6.51    & \nodata \\
            & & X & 2860    & 0.20   \times 0.16   & 66.8   & 2.66     & 5.00    & \nodata
 \setcounter{magicrownumbers}{0}
\enddata
\tablecomments{\textbf{Column 1.} target name. \textbf{Column 2.} frequency band observed. \textbf{Column 3.} observing time on target source. \textbf{Columns 4 \& 5.} restoring beam major axis, minor axis, and position angle for MTMFS images. \textbf{Column 6.} theoretical rms noise across band. \textbf{Column 7.} rms noise measured from MTFMS imaging results. \textbf{Column 8.} self-calibration technique applied (p: phase; a: amplitude). }
\label{tab:sensitivity}
\end{deluxetable*}

\subsection{Image Processing}
In Figures \ref{fig:vlatargets_C} and \ref{fig:vlatargets_X}, we show the multiterm, multiscale, multifrequency synthesis imaging results \citep[MTMFS;][also see \citetalias{2024ApJ...961..230S}]{2011A&A...532A..71R} spanning all calibrated data per band centered at 6 GHz and 10 GHz. For a majority of the images, we used two Taylor terms, multiple scales set to 0, 5, 10, 20, and 40 times the restoring beam, and Briggs deconvolution with a robust weighting of 0. Due to its complex structure, we imaged NGC~1068 with four Taylor terms for both bands to increase the achievable dynamic range.

We iteratively conducted self-calibration on any targets that initially showed imaging artifacts indicative of the instrumental response for the VLA. For each iteration we lightly cleaned an image to generate an approximate model of the source, then applied phase-only gain corrections with increasingly smaller solution intervals until the phases converged. If there were still artifacts, we then applied amplitude based corrections in the same manner. After self-calibration, NGC~1052, NGC~1068, NGC~3079, and NGC~5506 still showed imaging artifacts due to their higher brightness levels.\footnote{We attempted to use the wideband AW-projection gridder with conjugate beams and a single W-plane \citep[see][]{2013ApJ...770...91B}, but we found that these images resulted in a loss of faint emission.} These sources are dynamic range-limited due to non-closing baseline-based offsets, which is correctable only with significant observing time of a bright calibrator source (see VLA Scientific Memo \#152 by R.C. Walker).\footnote{The dynamic range limit for the VLA is ${\sim}10,000$ with standard calibration. The current record holder is by \citet[][]{2011A&A...527A.108S} using 3C~147 for a dynamic range of 1.6 million.}

Despite these dynamic range limitations, the artifacts in the images do not affect our science goals for this paper, since we are focused on the bright AGN in the center of each image. In Table \ref{tab:sensitivity} we report the restoring beam, the theoretical and measured rms noise values, and whether self-calibration was applied for each image. Note that no C-band observation for NGC~3786 occurred, and that the X-band observation of NGC~4180 failed due to poor weather.

We used \texttt{imfit} on each image with a single 2D Gaussian component to measure the peak flux density of the AGN within a region that was half of the maximum flux density component. We assumed that the AGN was within the point (or nearly point-like) sources detected in the majority of our imaging results. Similarly for sources with extended features, we assumed the AGN was within the bright point-like object near the pointing center. NGC~1068 contained multiple bright components, and the AGN is south of the brightest peak \citep{2023ApJ...953...87F}. We did not make a measurement of the nuclear region for this target due to the complex extended structure blending into the AGN, which prevented a direct measurement of the peak flux density for our imaging results. We report all the peak flux density measurements in Table \ref{tab:measurements}. All uncertainties include the uncertainty on the fit as determined by \texttt{imfit}, the rms noise of the cleaned image, and a 5\% systematic error (as is typical for the absolute flux calibration of the VLA) added in quadrature.

We also generated image cubes with a typical channel width of 128 MHz. NGC~3089 and NGC~7378 were faint, so we imaged each 2~GHz baseband individually with MTMFS using two Taylor terms. To obtain a consistent flux density measurement for comparison across the channels, we utilized a common $(u,v)$-taper of $0\farcs25$ and subsequently convolved each channel to the native beam of the lowest frequency image channel in the C band using \texttt{imsmooth}. In the X-band image cubes, we rotated the convolving beam to the position angle of the of the lowest frequency image channel.

\begin{figure*}[ht!]
    \centering
    \includegraphics[width=\textwidth]{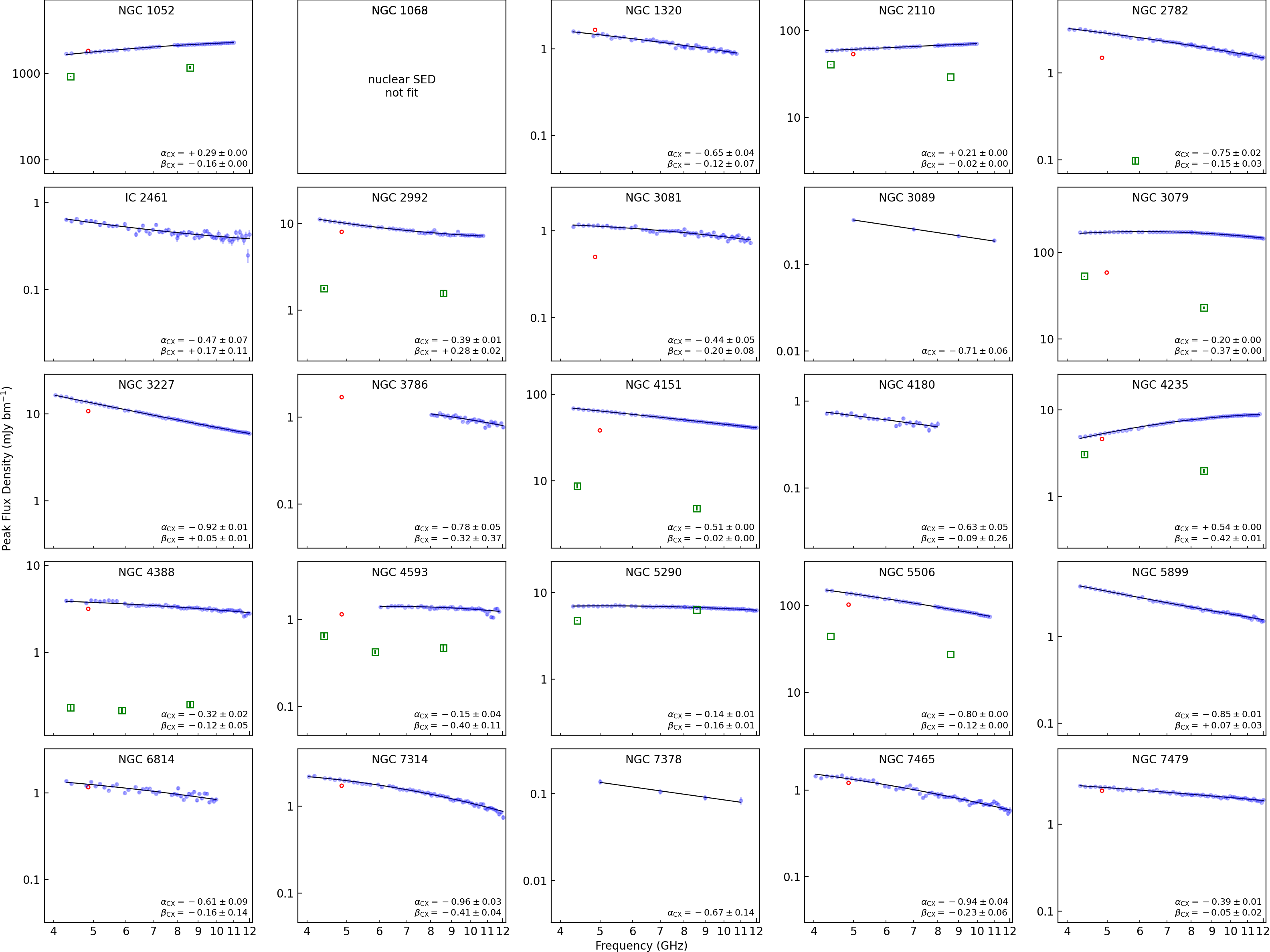}
    \caption{Spectral energy distributions for the VLA FRAMEx sample across $4-12$ GHz. Each plot spans two orders of magnitude on the vertical axis. The blue circles represent the peak flux density of the AGN from a 2D Gaussian fit on a single channel, where one channel is equal to a single 128 MHz spectral window for all targets except NGC 3089 and NGC 7378 where the channel width is from the 2-GHz baseband. The black line represents the spectral fitting results across $4-12$ GHz, using a power-law function with spectral curvature and the reference frequency centered at 8 GHz. The red circles represent the archival VLA peak flux densities listed in Table 4 of \citetalias{2021ApJ...906...88F} (4.9 GHz). The green squares represent the peak flux densities for VLBA detected targets from \citetalias{2022ApJ...936...76S} (5.9 GHz) and \citetalias{2024ApJ...961..109S} (4.4 GHz and 8.6 GHz). See \citetalias{2024ApJ...961..109S} for multifrequency VLBA fitting results.}
    \label{fig:vlatargets_sed}
\end{figure*}

\subsection{Spectral Energy Distributions}
In Figure \ref{fig:vlatargets_sed}, we present VLA SEDs of the AGNs across $4-12$ GHz using peak flux density measurements alongside their respective VLBA measurements from the 6 GHz (deep observation) imaging in \citetalias{2022ApJ...936...76S} and the multifrequency imaging at 4.4 and 8.6 GHz in \citetalias{2024ApJ...961..109S}, when available. We used \texttt{imfit} to measure the peak flux densities of the AGNs for each image channel using the C-band region we previously defined where the image channel rms noise was used as the uncertainty on the peak flux density. We then used these measurements to fit a spectral index across C band and X band, individually. For several targets we found that there appeared to be a systematic offset at the 8 GHz boundary between C band and X band. Many of the measured offsets between the two bands were within the assumed 5\% VLA systematic calibration error, particularly for data calibrated with 3C~286 as the primary flux calibrator. Larger offsets were measured for the brightest source (NGC~1052) and the faintest sources (typically with peak flux densities ${\lesssim}2~{\rm mJy/beam}$). A particularly large offset was measured for NGC~6814, where we measured a $31\%$ offset. It is not entirely clear why the difference between the bands was so large, but both of the C- and X-band observations for this target suffered from significantly more RFI than our other observations. The higher than normal RFI may potentially affect the calibration and compound with the faintness of the source to affect the flux scaling of the source. Thus for the FRAMEx sample we measured the spectral index with curvature across the entire $4-12$ GHz with respect to a reference frequency of 8 GHz while simultaneously fitting for the corrective offset.\footnote{We also attempted fitting with the reference frequency as a free parameter, but this led to poor constraints on the fit parameters.} 

In Table \ref{tab:spix}, we show the spectral index values for C band and X band respectively, as well as the spectral index with curvature across the full frequency range after X-band offset corrections were applied. Note that three targets in our sample were not fit for curvature: NGC~1068 lacked a clear nuclear peak so we did not do any spectral fitting, and both NGC~3089 and NGC~7378 were limited by their faintness and we only measured a spectral index with relatively few data points. The $\chi^2$ value and degrees of freedom for each fit along with the fitted peak flux density at the respective reference frequencies are listed in Appendix~\ref{apx:fits}. We found that the data were increasingly underfit as a function of source brightness, and this may be due to some unaccounted for in-band systematic uncertainty.\footnote{Though we did not include this in our final fitting results, we found $\frac{\chi^2}{\mathrm{dof}}\approx1$ when we included an in-band systematic uncertainty of $0.5\%\times S_{\rm peak}$ in the error for each image channel.}

\begin{deluxetable}{rrCC}
\tablecaption{AGN Peak Flux Densities}
\tablehead{
&
\colhead{Target} &
\colhead{$S_{\rm C}$} &
\colhead{$S_{\rm X}$} 
\\[-0.1cm]
& &
\colhead{(${\rm mJy~bm^{-1}}$)} &
\colhead{(${\rm mJy~bm^{-1}}$)}
}
\setcounter{magicrownumbers}{-25}
\startdata
\rowcolor{lightgray}  \rownumber & NGC~1052 & 1881   \pm   94  & 2015   \pm  101 \\
\rowcolor{lightgray} \rownumber & NGC~1068 & \nodata                & \nodata  \\
                     \rownumber & NGC~1320 &    1.10   \pm   0.06   &    0.84   \pm    0.04  \\
\rowcolor{lightgray} \rownumber & NGC~2110 &   60.1   \pm   3.0  &   68.8   \pm    3.4  \\
\rowcolor{lightgray} \rownumber & NGC~2782 &    1.56   \pm   0.08   &    0.80   \pm    0.05  \\
                     \rownumber &  IC~2461 &    0.46   \pm   0.02   &    0.32   \pm    0.02  \\
\rowcolor{lightgray} \rownumber & NGC~2992 &    60.4  \pm   3.02   &   69.4   \pm    3.47  \\
                     \rownumber & NGC~3081 &    0.85   \pm   0.04   &    0.65   \pm    0.03  \\
                     \rownumber & NGC~3089 &    0.22   \pm   0.01   &    0.14   \pm    0.01  \\
\rowcolor{lightgray} \rownumber & NGC~3079 &  167   \pm   8   &  157   \pm    8  \\
                     \rownumber & NGC~3227 &    7.89   \pm   0.44   &    3.75   \pm    0.20  \\
                     \rownumber & NGC~3786 & \nodata                &    0.58   \pm    0.03  \\
\rowcolor{lightgray} \rownumber & NGC~4151 &   53.4   \pm   2.67   &   39.7   \pm    1.99  \\
                     \rownumber & NGC~4180 &    0.45   \pm   0.02   & \nodata                \\
\rowcolor{lightgray} \rownumber & NGC~4235 &    6.14   \pm   0.31   &    8.63   \pm    0.43  \\
\rowcolor{lightgray} \rownumber & NGC~4388 &    2.79   \pm   0.14   &    2.23   \pm    0.11  \\
\rowcolor{lightgray} \rownumber & NGC~4593 &    1.36   \pm   0.07   &    1.27   \pm    0.06  \\
\rowcolor{lightgray} \rownumber & NGC~5290 &    6.85   \pm   0.34   &    6.64   \pm    0.33  \\
\rowcolor{lightgray} \rownumber & NGC~5506 &  113.0   \pm   5.7   &   74.4   \pm    3.7  \\
                     \rownumber & NGC~5899 &    2.05   \pm   0.11   &    0.86   \pm    0.06  \\
                     \rownumber & NGC~6814 &    1.05   \pm   0.05   &    0.43   \pm    0.03  \\
                     \rownumber & NGC~7314 &    1.61   \pm   0.08   &    0.72   \pm    0.04  \\
                     \rownumber & NGC~7378 &    0.11   \pm   0.01   &    0.07   \pm    0.01  \\
                     \rownumber & NGC~7465 &    0.77   \pm   0.04   &    0.38   \pm    0.02  \\
                     \rownumber & NGC~7479 &    2.35   \pm   0.12   &    2.07   \pm    0.10  
\enddata
\setcounter{magicrownumbers}{0}
\tablecomments{6 GHz and 10 GHz AGN peak flux densities as measured from the MTMFS imaging results shown in Figures \ref{fig:vlatargets_C} and \ref{fig:vlatargets_X}. The uncertainties include the 2D Gaussian pixel fitting error, the image rms, and a 5\% systematic uncertainty added in quadrature. The highlighted rows indicate VLBA-detected targets.}
\label{tab:measurements}
\end{deluxetable}

\begin{figure}[ht!]
    \centering
    \subfigure[]{\label{fig:flux_dist}\includegraphics[width=\columnwidth]{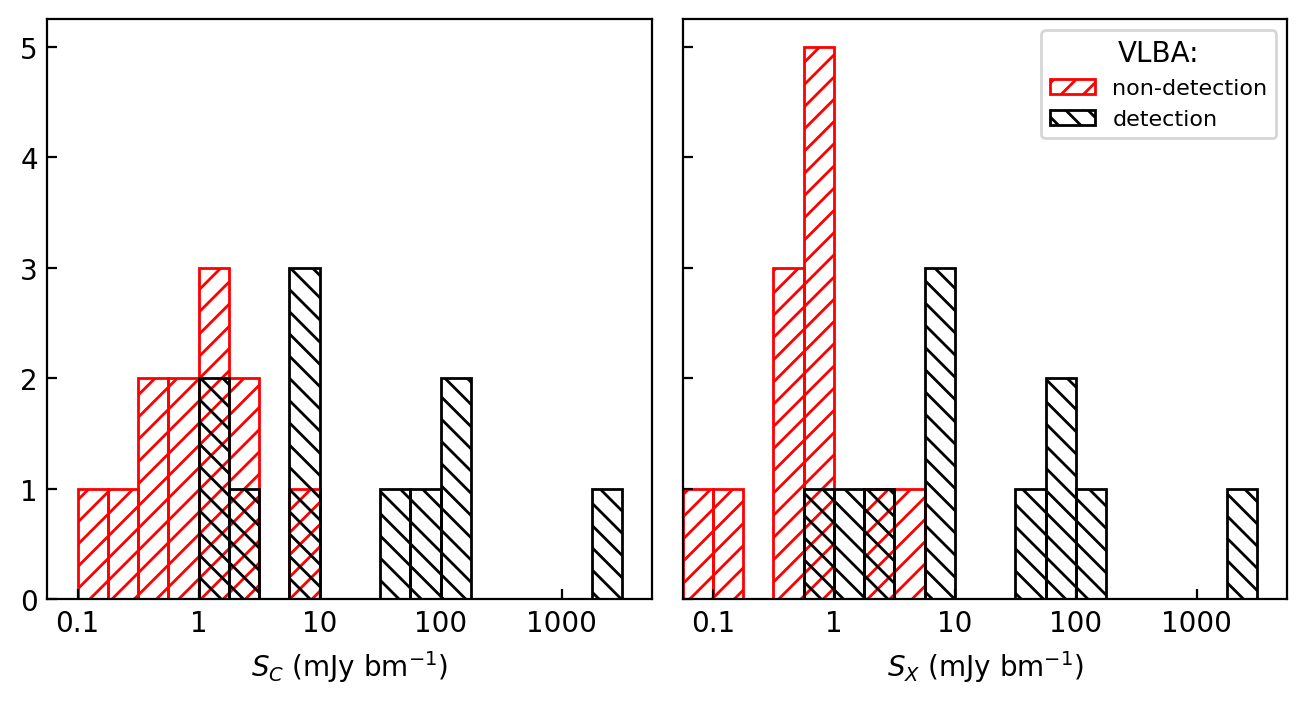}}\\
    \subfigure[]{\label{fig:sed_dist}\includegraphics[width=\columnwidth]{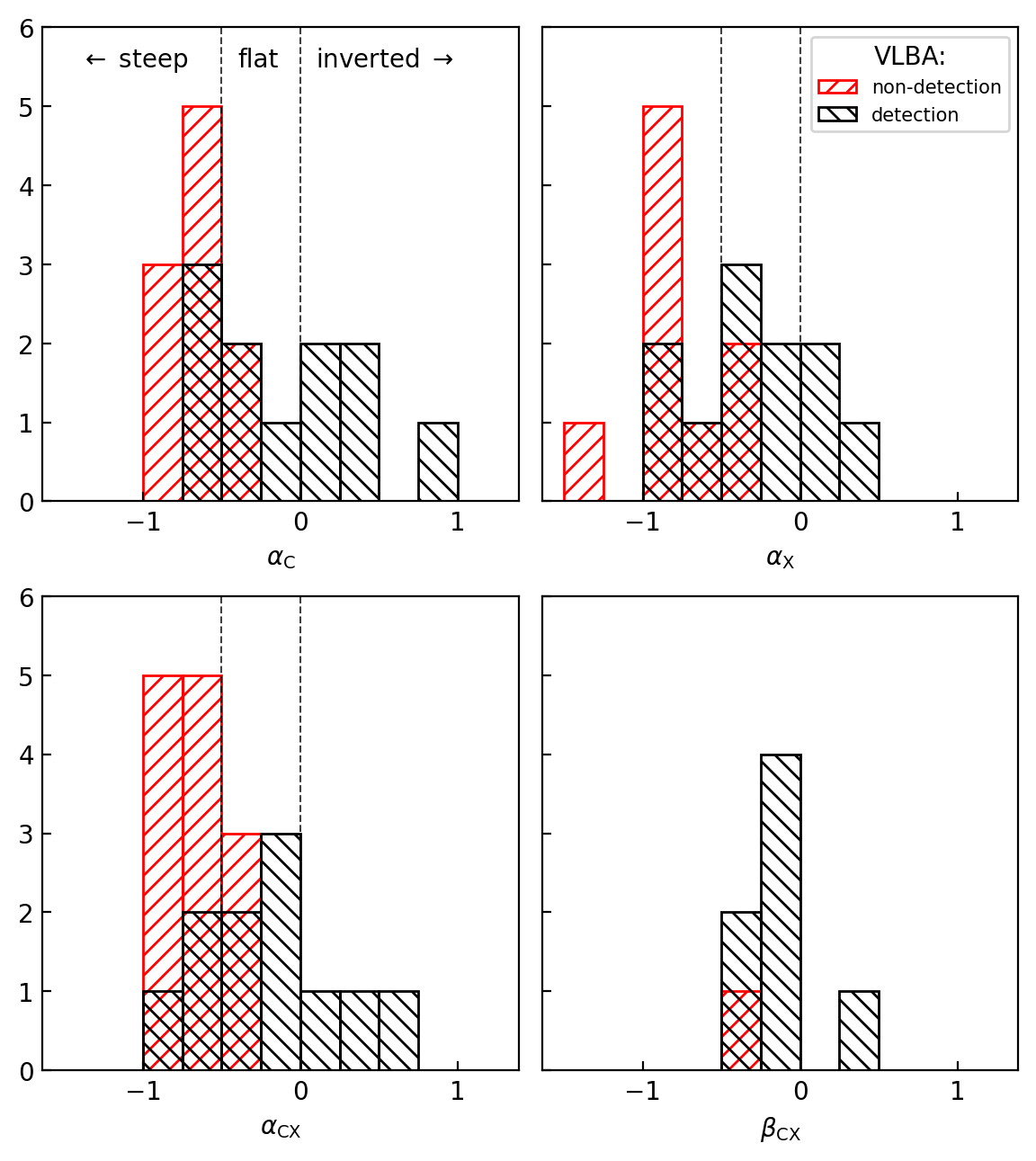}}
    \caption{Histograms of: \subref{fig:flux_dist} 6 GHz peak flux densities (left) and 10 GHz peak flux densities (right) from Table \ref{tab:measurements} as measured by the VLA; \subref{fig:sed_dist} Corresponding VLA $4-8$ GHz spectral indices (top left), $8-12$ GHz spectral indices (top right), $4-12$ GHz spectral indices (bottom left), and $4-12$ GHz spectral curvature values with significance (bottom right; see Table \ref{tab:spix}).}
    \label{fig:vlatargets_sed_distro}
\end{figure}

\begin{deluxetable*}{rrCCCCC}
\tablecaption{AGN Spectral Results}
\tablehead{
&
\colhead{Target} &
\colhead{$\alpha_{\rm C}$} &
\colhead{$\alpha_{\rm X}$} &
\colhead{$\alpha_{\rm CX}$} &
\colhead{$\beta_{\rm CX}$} &
\colhead{Corr.}
\\[-0.1cm]
\colhead{} &
\colhead{(1)} &
\colhead{(2)} &
\colhead{(3)} &
\colhead{(4)} &
\colhead{(5)} &
\colhead{(6)}
}
\setcounter{magicrownumbers}{-25}
\startdata
\rowcolor{lightgray} \rownumber & NGC~1052 &    0.39   \pm   ({<}0.01)  &    0.22  \pm   ({<}0.01) &    0.29  \pm    ({<}0.01)       &    \mathbf{ -0.16   \pm  ({<}0.01)}   &   +9.2   \pm  ({<}0.1)  \%  \\
\rowcolor{lightgray} \rownumber & NGC~1068 &    \nodata                 &  \nodata                 &  \nodata                        &    \nodata                            &      \nodata                 \\
                     \rownumber & NGC~1320 &   -0.58   \pm   0.02       &   -0.66  \pm   0.07      &   -0.65  \pm    0.04            &    -0.12   \pm   0.07                 &   -8.2   \pm   1.7       \%  \\
\rowcolor{lightgray} \rownumber & NGC~2110 &    0.22   \pm   ({<}0.01)  &    0.24  \pm   ({<}0.01) &    0.21  \pm    ({<}0.01)       &     -0.02  \pm   ({<}0.01)             &   -0.7   \pm   0.1       \%  \\
\rowcolor{lightgray} \rownumber & NGC~2782 &   -0.64   \pm   0.01       &   -0.83  \pm   0.03      &   -0.75  \pm    0.02            &    \mathbf{ -0.15  \pm   0.03  }      &   +4.3   \pm   1.0       \%  \\
                     \rownumber &  IC~2461 &   -0.59   \pm   0.05       &   -0.38  \pm   0.11      &   -0.47  \pm    0.07            &     0.17  \pm   0.11                  &   -9.4   \pm   3.3       \%  \\
\rowcolor{lightgray} \rownumber & NGC~2992 &   -0.58   \pm   0.01       &   -0.24  \pm   0.02      &   -0.39  \pm    0.01            &    \mathbf{ 0.26  \pm   0.02  }       &   -5.0   \pm   0.5       \%  \\
                     \rownumber & NGC~3081 &   -0.32   \pm   0.03       &   -0.54  \pm   0.08      &   -0.45  \pm    0.05            &    -0.20  \pm   0.08                  &   +1.5   \pm   2.1       \%  \\
                     \rownumber & NGC~3089 &    \nodata                 &  \nodata                 &   -0.71  \pm    0.06^{*}        &    \nodata                            &   \nodata                    \\
\rowcolor{lightgray} \rownumber & NGC~3079 &    0.02   \pm   ({<}0.01)  &   -0.38  \pm   ({<}0.01) &   -0.20  \pm    ({<}0.01)       &    \mathbf{ -0.37  \pm   ({<}0.01)}   &   -2.5   \pm   ({<}0.1)  \%  \\
                     \rownumber & NGC~3227 &   -0.95   \pm   ({<}0.01)  &   -0.91  \pm   0.01      &   -0.92  \pm    0.01            &    0.05  \pm   0.01                   &   +2.0   \pm   0.3       \%  \\
                     \rownumber & NGC~3786 &    \nodata                 &   -0.76  \pm   0.04      &   -0.78  \pm    0.05^{\dagger}  &    -0.32  \pm   0.37^{\dagger}        &   \nodata                    \\
\rowcolor{lightgray} \rownumber & NGC~4151 &   -0.50   \pm   ({<}0.01)  &   -0.50  \pm   ({<}0.01) &   -0.51  \pm    ({<}0.01)       &    \mathbf{-0.02  \pm   ({<}0.01)}    &   -0.7   \pm   ({<}0.1)  \%  \\
                     \rownumber & NGC~4180 &   -0.62   \pm   0.04       &  \nodata                 &   -0.63  \pm    0.05^{\ddagger} &    -0.09  \pm   0.26^{\ddagger}       &   \nodata                    \\
\rowcolor{lightgray} \rownumber & NGC~4235 &    0.79   \pm   ({<}0.01)  &    0.36  \pm   0.01      &    0.54  \pm    ({<}0.01)       &    \mathbf{ -0.42  \pm   0.01  }      &   -3.5   \pm   0.2       \%  \\ 
\rowcolor{lightgray} \rownumber & NGC~4388 &   -0.25   \pm   0.03       &   -0.37  \pm   0.03      &   -0.32  \pm    0.02            &    -0.12  \pm   0.05                  &   +2.7   \pm   0.9       \%  \\
\rowcolor{lightgray} \rownumber & NGC~4593 &   -0.08   \pm   0.07       &   -0.28  \pm   0.04      &   -0.15  \pm    0.04            &    -0.40  \pm   0.11                  &   -1.6   \pm   1.3       \%  \\
\rowcolor{lightgray} \rownumber & NGC~5290 &   -0.05   \pm   0.01       &   -0.21  \pm   0.01      &   -0.14  \pm    0.01            &    \mathbf{  -0.16  \pm   0.01  }     &   -2.5   \pm   0.3       \%  \\
\rowcolor{lightgray} \rownumber & NGC~5506 &   -0.71   \pm   ({<}0.01)  &   -0.85  \pm   ({<}0.01) &   -0.80  \pm    ({<}0.01)       &    \mathbf{  -0.12  \pm   ({<}0.01) } &   -2.7   \pm   ({<}0.1)  \%  \\
                     \rownumber & NGC~5899 &   -0.90   \pm   0.01       &   -0.82  \pm   0.02      &   -0.86  \pm    0.02            &    0.07  \pm   0.03                   &   -1.7   \pm   0.7       \%  \\
                     \rownumber & NGC~6814 &   -0.51   \pm   0.03       &   -0.66  \pm   0.19      &   -0.61  \pm    0.09            &    -0.16 \pm 0.14                     &  +31.0   \pm   2.1       \%  \\
                     \rownumber & NGC~7314 &   -0.68   \pm   0.01       &   -1.21  \pm   0.05      &   -0.96  \pm    0.03            &    \mathbf{  -0.41  \pm   0.04 }      &  +13.3   \pm   1.1       \%  \\
                     \rownumber & NGC~7378 &    \nodata                 &  \nodata                 &   -0.67  \pm    0.14^{*}        &    \nodata                            &   \nodata                    \\
                     \rownumber & NGC~7465 &   -0.79   \pm   0.02       &   -0.90  \pm   0.06      &   -0.94  \pm    0.04            &    -0.23  \pm   0.06                  &   -7.0   \pm   1.8       \%  \\
                     \rownumber & NGC~7479 &   -0.37   \pm   0.01       &   -0.40  \pm   0.02      &   -0.39  \pm    0.01            &    -0.05  \pm   0.02                  &  -10.8   \pm   0.7       \% 
\enddata
\tablecomments{Spectral fitting results with formal uncertainties for peak flux densities from channelized data for the VLA FRAMEx sample. Highlighted rows indicate VLBA-detected sources. \textbf{Column 1.} target name. \textbf{Column 2.} $4-8$ GHz spectral index. \textbf{Column 3.} $8-12$ GHz spectral index \textbf{Column 4.} $4-12$ GHz spectral index at 8 GHz. \textbf{Column 5.} $4-12$ GHz spectral curvature at 8 GHz. Boldface values indicate ``significant'' curvature, i.e. sources with ${\rm abs}(\beta/\sigma_{\beta})>3$ that also pass a nested F-test with 99.7\% confidence. \textbf{Column 6.} fitted percent shift of X-band data to correct for systematic offset between C band and X band. \\
$^{*}$NGC~3089 and NGC~7378 were not fit with curvature due to their faintness.\\
$^{\dagger}$NGC~3786 was fit with a reference frequency set to 10 GHz since there were only X-band observations.\\
$^{\ddagger}$NGC~4180 was fit with a reference frequency set to 6 GHz since there were only C-band observations.
}
\label{tab:spix}
\end{deluxetable*}

\section{Results} \label{sec:results}

\subsection{Brightness Properties}

In Figure \ref{fig:flux_dist} we show the distribution of peak flux densities of the AGNs from Table \ref{tab:measurements} and we distinguish by whether the target was detected by the VLBA or not. For the VLBA-detected objects, the faintest targets in these VLA observations are NGC~2782, NGC~4388, and NGC~4593, with peak flux densities approximately between $1-3~{\rm mJy~bm^{-1}}$. These also happen to be the only three targets that were detected by the 4-hour VLBA deep observations in \citetalias{2022ApJ...936...76S}, which probed a sensitivity limit of $\sim10~\rm{\mu Jy~bm^{-1}}$. The remainder of the VLBA-detected objects all have VLA peak flux densities ranging between $6-170~{\rm mJy~bm^{-1}}$ except the radio-loud target NGC~1052, which has a peak flux density on the order of $2~{\rm Jy~bm^{-1}}$.

For the objects not detected by the VLBA, we note that all have VLA peak flux densities ${<}8~{\rm mJy~bm^{-1}}$. NGC~1320, NGC~3081, NGC~3089, NGC~6814, NGC~7314, and NGC~7465 were observed but not detected in the VLBA deep observations and have VLA peak flux densities ranging between $0.22-1.61~{\rm mJy~bm^{-1}}$ in the C band and $0.14-0.84~{\rm mJy~bm^{-1}}$ in the X band. The remaining seven objects (IC~2461, NGC~3227, NGC~3786, NGC~4180, NGC~5899, NGC~7378, NGC~7479) have VLA peak flux densities ranging between $0.11-7.89~{\rm mJy~bm^{-1}}$ in the C band and $0.07-3.75~{\rm mJy~bm^{-1}}$ in the X band. The deep observation sample was selected from the archival VLA observations with point-like morphology \citepalias[shown in Figure 3 of ][]{2021ApJ...906...88F} for a higher likelihood of detection with the VLBA. NGC~7479 was not re-observed in \citetalias{2022ApJ...936...76S} and may be a potential candidate for a deep observation, as it has a similar brightness levels to the three detected objects in the deep observations. While NGC~3227 and NGC~5899 appear to have some extended morphology, the peak flux density of the AGNs is also similar to that of the deep observation detections and may be possible candidates for re-observation. The remaining four objects have VLA peak flux densities ${<}1~{\rm mJy~bm^{-1}}$ and may be less likely to have detections at VLBA spatial scales.

\subsection{Spectral Properties}

\begin{figure}[ht!]
    \centering
    \includegraphics[width=\columnwidth]{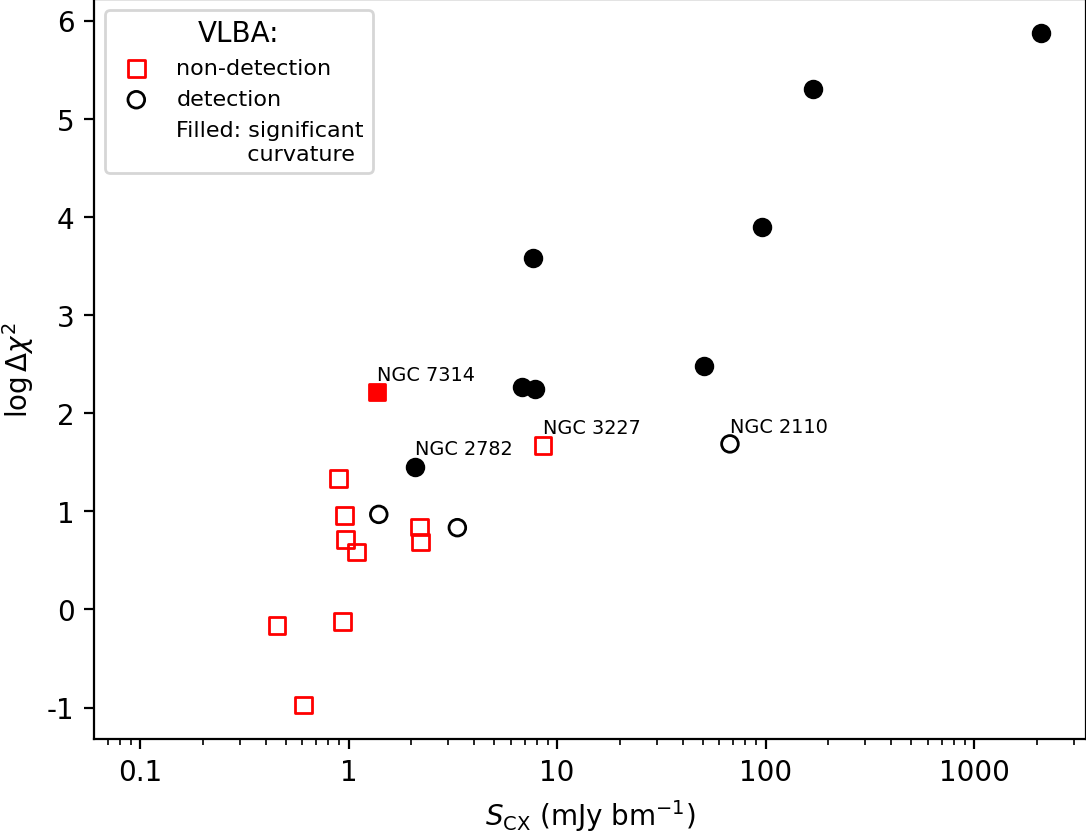}
    \caption{Difference in $\chi^2$ values between power-law model fits with and without spectral curvature, plotted against source brightness at the reference frequency. Black circles represent VLBA-detected objects, while red squares denote non-VLBA-detected objects. Filled markers highlight sources with significant curvature, defined as $\rm{abs}(\beta/\sigma_{\beta})>3$ and rejecting the null hypothesis of no curvature at 99.7\% confidence.}
    \label{fig:ftest}
\end{figure}

In Figure \ref{fig:sed_dist} we show the distribution of spectral index and spectral curvature values from Table \ref{tab:spix}, and we have again distinguished whether the target was detected in FRAMEx VLBA observations or not. When using typical spectral classifications in which $\alpha<-0.5$ is considered steep, $-0.5\leq\alpha\leq0.0$ is flat, and $\alpha>0.0$ is inverted, we observe that there is a clear preference towards a flat or an inverted spectral index for the detected objects. Conversely, we find no inverted indices for the objects not detected by the VLBA, all of those indices are $-1.0\lesssim\alpha_{\rm CX}\lesssim-0.4$ and they generally tend towards a steep spectral index. 

In Table \ref{tab:spix}, we have indicated nine targets with significant spectral curvature which we also depict in the bottom right of Figure \ref{fig:sed_dist}. We define significant curvature as those meeting the criteria that $\mathrm{abs}(\beta/\sigma_{\beta})>3$ which also reject the null hypothesis that there is no spectral curvature with 99.7\% confidence in an F-test.
The statistical significance of spectral curvature for each source is depicted in Figure \ref{fig:ftest}, where we show the difference of $\chi^2$ values between the power-law and the power-law with curvature fits as a function of source brightness. There appears to be dependence on the brightness levels for whether significant curvature is detected since the brightest sources, primarily the VLBA detected objects, indicate significant curvature.
VLBA-detected targets with statistically significant curvature are NGC~1052, NGC~2782, NGC~2992, NGC~3079, NGC~4151, NGC~4235, NGC~5290, and NGC~5506. NGC~7314 is the only target with significant curvature that is not detected by the VLBA.
NGC~2110, NGC~4388, and NGC~4593 are the three VLBA-detected sources without significant curvature, but detection of intrinsic curvature in NGC~4388 and NGC~4593 may be limited in our observations by their faintness. We discuss observational implications of the detection of spectral curvature in the radio in Section \ref{sec:curvature}.

\section{Discussion} \label{sec:discussion}

\subsection{Observational Comparisons}
Out of our volume-complete sample of 25 hard X-ray selected AGNs, FRAMEx has thus far detected 12 with the VLBA. In \citetalias{2021ApJ...906...88F} we suggested that VLA measurements of the emission surrounding the AGNs were likely contaminated by extranuclear host interactions that largely resolve away when observed at the resolution of the VLBA. This is especially true for the sources not detected by the VLBA and the fainter sources detected only by the deep observations in \citetalias{2022ApJ...936...76S}, where we have recovered only $\sim4\%$ of the VLA detected emission in NGC~2782 and $\sim6\%$ in NGC~4388. Two classes of objects emerge from this work when the FRAMEx sources are separated by whether they were detected by the VLBA as depicted in Figure \ref{fig:vlatargets_sed_distro}. The VLBA detected objects have a clear preference towards the brightest VLA measurements with flat to inverted spectra while the non-detections are the faintest sources and have steep spectra. This could potentially indicate that different radio emission mechanisms are being probed, the measurement of which depending on whether there is a compact radio source.

The non-detected sources with the VLBA have a mean VLA spectral index of $\langle\alpha^{\rm ND}_{\rm CX}\rangle=-0.69$ (with a scatter of $\sigma_{\alpha^{\rm ND}_{\rm CX}}=0.18$), remarkably similar to spectral indices observed in star-forming galaxies \citep[SFGs; see, e.g.][]{1982A&A...116..164G,1983ApJS...53..459C,1989MNRAS.236..737E}, including the prototypical SFG M82 \citep[$\alpha=-0.8$;][]{1992ARA&A..30..575C}. The similarity to that of SFGs could hint that the bulk of the emission observed by the VLA is dominated by recent star-formation activity in \ion{H}{2} regions \citep{1992ARA&A..30..575C}. However, in \citetalias{2024ApJ...961..230S} we found that the star formation rate (SFR) from the radio emission surrounding the AGN in NGC~4388  was too luminous for star formation alone when compared to optical tracers, even though the VLBA emission is only $\sim6\%$ of the VLA emission. That analysis led us to suggest that AGN winds were a plausible source for the excess radio emission in the ISM surrounding the AGN.

Though a full study of SFRs for the FRAMEx sample is beyond the scope of this work,  a comparison to hard X-ray selected studies of AGNs may provide some initial clues as to what the origins of radio emission is at VLA resolution. \cite{2022MNRAS.515..473P} observed a hard X-ray selected sample ($20-40~\rm{keV}$) of 30 AGNs with the VLA to measure matched $\sim1\arcsec$ resolution SEDs at several frequencies between 5 and 45 GHz. Their sample differs from ours by extending to further distances ($\sim14-1700~{\rm Mpc}$) than the $40~{\rm Mpc}$ FRAMEx limit, and it includes some high brightness ``radio-loud'' AGNs that they define as sources with clear signatures of relativistic jets. The remainder of their targets are considered ``radio-quiet'' and may be most similar to the FRAMEx sample, as NGC~1052 is the only object exhibiting a clear jet-like structure (spanning $\sim27''$, or $\sim2.8~\rm{kpc}$, and terminating in radio lobes in our C-band imaging results). They suggest that the radio quiet AGNs consist of several categories based on their SEDs, with 37\% of their sources exhibiting steep spectra from unresolved optically thin synchrotron emission (e.g., from subrelativistic jets, star formation, winds), 10\% of their sources having GHz-peaked-spectra (see next section in this work), and the remaining sources having flat spectra compatible with optically thick synchrotron emission (e.g., from a compact jet or magnetically heated corona). Though we do not have as broad of an SED as the \cite{2022MNRAS.515..473P} work, it is possible that our sample may have similar statistics, which may indicate that the flat spectrum VLBA-detected sources in the FRAMEx sample are due to radio emission from an unresolved jet or an optically thick coronal source.

\cite{2022MNRAS.515..473P} go on to suggest following-up with VLBI observations, as the high-resolution observations may help to further constrain the radio emission origins by ruling out processes such as star formation and AGN winds. The FRAMEx project is unique when considering that we now have a uniform set of VLA measurements which are complemented by quasi-simultaneous observations of the same targets with the VLBA and the \textit{Swift} XRT. Our VLA results crossmatched with our VLBA observational detections have clearly distinguished at least two classes of AGN based on their spectra and brightness levels. We will further investigate the consequences of the compact radio source detections (or not) by exploring the potential radio emission mechanisms of our sample in detail in Paper~III.

\subsection{Spectral Curvature}

Spectral curvature can occur at frequencies near the transition between for instance optically thin and optically thick synchrotron emission, but also other emission processes.
Detailed spectral curvature measurements in the radio are limited due to bandwidth constraints or the typically static antenna configurations for interferometers, as specified in Section \ref{sec:intro}. In many cases, spectral curvature is roughly estimated from measurements at three or more frequencies \cite[e.g.,][]{1987ApJ...313..651E,2022MNRAS.512..471C, 2024ApJ...961..109S}. Measurements with high frequency resolution have only recently been made possible with the upgraded VLA, which now has a continuous $1-50$ GHz frequency coverage and can sample a maximum bandwidth of 8 GHz in each polarization.
Our VLA spectral measurements were performed for a given spectral range and array configuration and our results motivate gathering even broader spectra with matched spatial resolutions. However, it may take multiple observing semesters to fully take advantage of the VLA's antenna configurations in order to gather a matched resolution SED across a broad frequency range, prolonging the scientific analysis. The Next Generation VLA (ngVLA) could provide a solution: by utilizing the sub-arraying capabilities of the observatory, a target could be observed with a compact array at higher frequencies simultaneously with an extended array for the lower frequencies. This would provide the capability to measure an instantaneous full synchrotron spectrum over similar spatial scales, potentially providing evolutionary clues to the origins of the radio source as we discuss in Section \ref{sec:gps}.

\subsubsection{Intrinsic Curvature vs. Observational Limitations}
\label{sec:curvature}
We find that 9 out of the 22 spectral fits show statistically significant curvature in our VLA spectra, and Figure \ref{fig:ftest} shows that these targets coincide with the brightest sources in the sample. We therefore suggest that radio spectral curvature may be detectable for the fainter sources if there is sufficient signal-to-noise. For our sensitivity limits, the brightness significance threshold appears to be $\sim5~\rm{mJy~bm^{-1}}$, and in Figure \ref{fig:ftest} we have indicated four sources that straddle this threshold, which we discuss here briefly. NGC~2110 and NGC~3227 are both well above the $5~\rm{mJy~bm^{-1}}$ brightness threshold but are significantly not curved between $4-12$ GHz. NGC~3227 is the brightest of the VLBA non-detections and has a similar brightness level to three other targets with a significant curvature. We expect that any intrinsic curvature between $4-12$ GHz would be apparent, yet this source appears to have a power-law spectrum without any curvature. The lack of VLBA detection for this source combined with having one of the steepest spectral indices of the sample ($\alpha=-0.92\pm0.01$) perhaps indicates that radio emission in the target may be dominated by star formation. On the contrary, NGC~2110 is one of the brightest sources in our sample and has an inverted VLA spectrum that appears to be intrinsically not curved at these frequencies, suggesting mechanisms which dominate in the radio over pure star forming processes. Below the $5~\rm{mJy~bm^{-1}}$ threshold, NGC~7314 is the faintest target with significant spectral curvature and is one of the most curved of our sample at $\beta=-0.41\pm0.04$, indicating that strong intrinsic curvature can be apparent in weakly emitting sources. Lastly, NGC~2782 has the lowest significance of the entire sample and is one of the three detected targets in the \citetalias{2022ApJ...936...76S} VLBA deep observations (NGC~4388 and NGC~4593 being the other two), but is the only one with statistically significant curvature.

\subsubsection{Evolutionary Tracks}
\label{sec:gps}

The synchrotron spectrum of a statistically representative sample of sources such as FRAMEx may be particularly useful in understanding their evolutionary tracks, as was discussed in Section 4.4 of \citetalias{2024ApJ...961..109S}. The multifrequency flux density measurements in \citetalias{2024ApJ...961..109S}  suggested a turnover frequency in several sources comparable to that of GHz-peaked spectrum (GPS) sources, which are thought to be compact, young radio sources that peak in flux density $\sim 0.5-5~{\rm GHz}$, above which their spectra become steep \citep{2014MNRAS.438..463O,2021A&ARv..29....3O}. GPS sources are part of a family of related sources that includes high frequency peakers (HFPs), which peak above 5 GHz, and the compact steep spectrum sources (CSS), which extend from $\sim 1-20~{\rm kpc}$ and tend to peak below $\sim0.5$ GHz \citep[for a recent review see][and references therein]{2021A&ARv..29....3O}. The peak frequencies in these objects are anti-correlated with the angular size of their host galaxies ($\nu_{\rm pk}\propto\theta^{-4/5}$) and this may suggest that these objects follow an evolutionary track where the compact and young HFPs evolve into GPS sources, then to CSS objects, and finally into large scale radio galaxies.

\cite{2014MNRAS.438..463O} measured SEDs for four HFPs with the VLBA and showed that these sources have increasingly resolved morphology, exhibiting single components at low frequencies and resolve into multiple components at higher frequencies. They stacked each sub-component's respective SEDs at VLBA resolution, and found that the stacked VLBA SEDs characterized the corresponding VLA SEDs quite well: when compared to the previous VLA observations \citep{2000A&A...363..887D,2005A&A...432...31T,2007A&A...475..813O}, the spectral peaks observed with the VLBA contained $\gtrsim90\%$ of the corresponding flux density observed by VLA, and only a marginal  shift in the peak frequency ($\lesssim1.5~{\rm GHz}$). This suggests that much of the emission detected at VLBA scales is simply unresolved in the VLA observations for HFPs, indicating a divergence from the FRAMEx targets where some sources have emission measured by the VLBA that is $<10\%$ of the VLA (see Figure \ref{fig:vlatargets_sed}).

There are also differences in the FRAMEx targets when comparing the VLA and VLBA SEDs, in contrast to the previously discussed HFPs. Our VLA results suggest that NGC~1052 clearly peaks above the 12 GHz limiting frequency, but the \citetalias{2024ApJ...961..109S} VLBA results suggest that the nuclear spectrum peaks around $\nu_{\rm pk}=5.3\pm1.7~{\rm GHz}$. Similarly, NGC~4235 is a VLA source with curvature that peaks at $\nu_{\rm pk}=4.3\pm1.2~{\rm GHz}$ in the VLBA measurements but the VLA spectrum is inverted at the VLBA peak. Furthermore, NGC~2110 and NGC~4151 have little to no curvature but clearly peak at the VLBA resolution, $\nu_{\rm pk}=4.9\pm1.7~{\rm GHz}$ and $\nu_{\rm pk}=2.9\pm1.1~{\rm GHz}$, respectively. NGC~2992 is another VLBA-detected object with significant VLA curvature (and the only one with positive curvature), but it has a flat spectrum at VLBA resolution. The fact that the VLA spectra differ quite drastically from the VLBA ones may indicate that the FRAMEx sources are not within the same family as GPS, HFPs, and CSS type sources. This may suggest that the emission at VLA resolution when compared to the VLBA emission is indeed extranuclear and we model this excess radio emission in NGC~3079 in the next section.

\begin{figure}[ht!]
    \centering
    \includegraphics[width=\columnwidth]{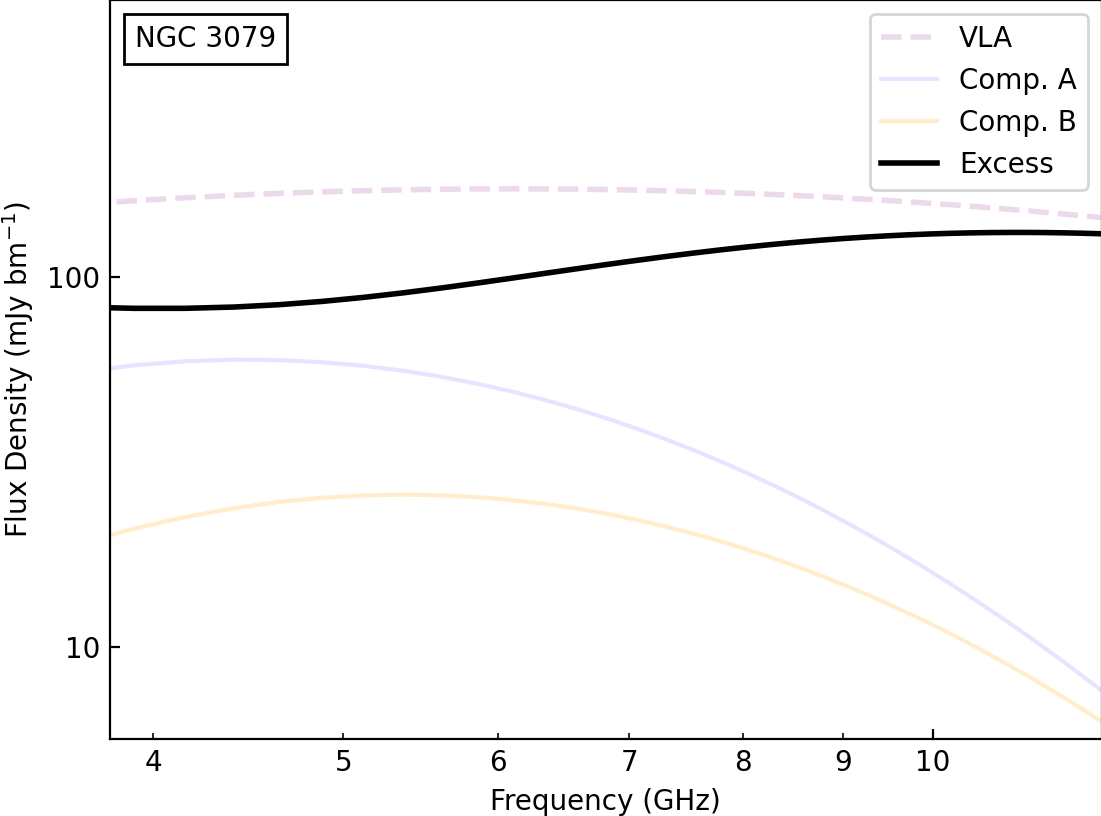}
    \caption{SEDs across $4-12~{\rm GHz}$ for emission from various components in NGC~3079. The dashed line represents the fit from the VLA SED from Table \ref{tab:spix}. The colored solid lines represent the fitted SEDs for the two sub-components from the VLBA measurements in \citetalias{2024ApJ...961..109S}. The solid black line represents the SED of the excess emission that is detected at VLA resolution but not at VLBA resolution.}
    \label{fig:ngc3079}
\end{figure}

\subsection{The Substructure of the AGN in NGC~3079}
NGC~3079 is one of two targets in the FRAMEx sample that resolves into multiple compact radio components at VLBA resolution \citetalias{2023ApJ...958...61F}. This source shows significant curvature at VLA resolution, and there appears to be a spectral break near the 8 GHz frequency edge of the two bands:  $\alpha_{\rm C}=0.02\pm0.01$ and $\alpha_{\rm X}=-0.38\pm0.01$. However, it is not a priori clear whether this is representative of a real spectral break or if it is due to a systematic difference between the two observations for NGC~3079. Given the brightness of NGC~3079, we attempted to calibrate it without phase referencing, to see if there were any effects from bootstrapping the flux densities from the phase calibrator source, but we measured similar indices. 

It is highly unlikely that any variability is affecting the measurements, as our two VLA observations for this object occurred consecutively. It is possible however that the measured spectral shape is due to the underlying structure of the sources not resolvable by the VLA. There are two bright radio knots (Components A and B) surrounding the AGN  at VLBI resolution \citep{1988ApJ...335..658I,1998ApJ...495..740T,2000PASJ...52..421S}. Both bright components have been observed to vary, and Component A has continuously increased in brightness since 1992 \citep[see Figure 3 in][]{2023ApJ...958...61F}. Component B peaked in brightness around 2005 and has meanwhile steadily decreased in brightness. Additionally, the components are separating, and \citetalias{2023ApJ...958...61F} measured a constant projected separation rate of $(0.040\pm0.003)c$ for Components A and B since a slowdown event that occurred around 2004. 

The two components do appear to have differing spectral peaks at VLBA resolution, and in Figure \ref{fig:ngc3079} we show the SED as measured by the VLA in this work compared to the multiwavelength fit from the VLBI measurements in \citetalias{2024ApJ...961..109S}. Neither component was detected at 22 GHz in \citetalias{2024ApJ...961..109S}, in contrast to the $\sim 1''$ resolution, 22 GHz VLA observation (C-array configuration), where a peak flux density of $63~\rm{mJy~bm^{-1}}$ is measured (Magno et al. 2025, submitted). This implies that the VLBA spectra of both components peaks in the $4-6~{\rm GHz}$ range, perhaps similar to our VLA fitting results. We have therefore modeled the substructure for a third unknown excess emission component that is measured by the VLA but not the VLBA, and we show the resulting model in Figure \ref{fig:ngc3079}, assuming that the \citetalias{2024ApJ...961..109S} VLBA spectral structure is representative across the $4-12$ GHz band. The modeled radio structure implies that the excess contributor has an approximately flat spectrum, or is slightly increasing with frequency, with a brightness of ${\sim}100~{\rm mJy~bm^{-1}}$ (with the flat spectra perhaps continuing to 22 GHz at $1''$ resolution), and is brighter than both Components A and B at high resolution. The faint third component observable by the VLBA is not well constrained and has an insignificant contribution relative to the other components, and cannot account for the excess flux observed by the VLA. The modeled SED implies that in this scenario, the majority of the excess flux in this source must be emitted at distances larger than the resolving capability of the VLBA, i.e., $\gtrsim1~{\rm pc}$, but smaller than the VLA resolution, i.e., $\lesssim40~{\rm pc}$, particularly for higher frequencies. The VLBI measurements of NGC~3079 show that there is a complex source encapsulated entirely within the restoring beam of our VLA observations and we will further explore the excess emission in Paper III.

\section{Summary and Conclusions} \label{sec:conclusion}

We have measured in great detail the circumnuclear radio spectra across $4-12~{\rm GHz}$ with the VLA for a uniform sample of nearby AGNs, based on a volume-complete hard X-ray selection. The sample has also been observed at milli-arcsecond resolution with the VLBA. Our conclusions are as follows:

\begin{enumerate}
    \item Our sample distinguishes between two families of AGN emission when analyzing VLA spectra and sorting the results by whether the AGN was detected at VLBA resolution. The VLBA non-detected population have a mean spectral index value of $\langle\alpha^{\rm ND}_{\rm CX}\rangle=-0.69$ with a scatter of $\sigma_{\alpha}=0.18$, in line with optically thin synchrotron spectra, and measures extranuclear radio emission. The VLBA detected population trends toward flat or inverted spectra and measures a combination of extranuclear emission at VLA resolution and nuclear emission at VLBA resolution.
    \item We find that the brightest sources have statistically significant spectral curvature, including eight VLBA detected sources and one VLBA non-detected source, implying spectral curvature is detectable with high signal-to-noise. Two sources have no curvature with statistical significance including one VLBA-detected source, NGC~2110.
    \item The nonmatching flux densities and shifted spectral peaks at VLA vs VLBA resolution implies that the FRAMEx sample does not appear to have SEDs that correlate well with the spectral signatures observed for GHz peaked spectrum (GPS) sources. The spectral structure from VLA observations of the GPS objects appears to correlate strongly with their corresponding VLBA observations, further indicating that radio emission for the FRAMEx objects measured by the VLA is indeed extranuclear.
    \item We modeled the excess flux density measured by the VLA but not the VLBA for NGC~3079 and found that SED is flat or perhaps slightly inverted. The excess flux density measured by the VLA is on the order of $\sim 100~{\rm mJy~beam^{-1}}$ and is produced beyond the approximately parsec spatial scales observed by the VLBA.
\end{enumerate}

Future work in this series will focus on understanding the VLBA vs VLA emission, multiwavelength analysis, and polarization of each source.

\section*{Acknowledgements}

We thank the anonymous referee for the instructive comments. We thank Bryan Butler, Viral Parekh, Rick Perley, Lilia Tremou, and the NRAO Help Desk for their great support in this work. 

The National Radio Astronomy Observatory is a facility of the National Science Foundation operated under cooperative agreement by Associated Universities, Inc. The authors acknowledge use of the Very Long Baseline Array under the US Naval Observatory’s time allocation. This work supports USNO’s ongoing research into the celestial reference frame and geodesy.
\facilities{VLA, VLBA}
\software{CASA, astropy \citep{astropy:2022}, scipy \citep{2020SciPy-NMeth}}

\bibliography{sample631}{}
\bibliographystyle{aasjournal}

\appendix

\section{Additional SED Fitting Results}
\label{apx:fits}

Table \ref{tab:chi2fits} contains both the peak flux densities from the spectral fits at the reference frequencies and the $\chi^2$ values from each fit. For the $4-12$ GHz fitting results, we conducted an F-test:
\begin{equation}
    F(k,N-m-k)=\frac{\Delta {\rm RSS}/k}{{\rm RSS_2}/(N-m-k)}
\end{equation}
where $\Delta {\rm RSS}={\rm RSS}_1-{\rm RSS}_2$ is the difference between the residual sums of squares (i.e., the $\chi^2$ values) of the power law model (${\rm RSS_1}$) and the power law with curvature model (${\rm RSS_2}$), $N$ is the number of data points used in the fits,  $m$ is the number of parameters in the power law model ($m=2$), and $k$ is the number of extra parameters in fitting power law with curvature model ($k=1$).

\begin{rotatetable*}
\begin{deluxetable}{rrC|CC|CC|CCCC}
\centering
\tablecaption{AGN peak flux density and reduced $\chi^2$ from SED fits}
\tabletypesize{\footnotesize}
\tablehead{
&
\colhead{Target} &
\colhead{Beam} &
\colhead{$S_{\rm C,fit}$} &
\colhead{$\frac{\chi^2_{\rm C}}{\rm dof}$} &
\colhead{$S_{\rm X,fit}$} &
\colhead{$\frac{\chi^2_{\rm X}}{\rm dof}$} &
\colhead{$S_{\rm CX,fit}$} &
\colhead{$\frac{\chi^2_{\rm CX}}{\rm dof}$} &
\colhead{$\frac{\Delta{\rm RSS}}{N-m-k}$} &
\colhead{$p$-value}
\\[-0.1cm]
\colhead{} &
\colhead{} &
\colhead{($\rm arcsec\times arcsec$)} &
\colhead{(${\rm mJy~bm^{-1}}$)} &
\colhead{} &
\colhead{(${\rm mJy~bm^{-1}}$)} &
\colhead{} &
\colhead{(${\rm mJy~bm^{-1}}$)} &
\colhead{} &
\colhead{} &
\colhead{}
\\[-0.1cm]
\colhead{} &
\colhead{(1)} &
\colhead{(2)} &
\colhead{(3)} &
\colhead{(4)} &
\colhead{(5)} &
\colhead{(6)} &
\colhead{(7)} &
\colhead{(8)} &
\colhead{(9)} &
\colhead{(10)} }
\setcounter{magicrownumbers}{-25}
\startdata
\rowcolor{lightgray} \rownumber & NGC~1052 & 0.94 \times 0.51 & 1888.84 \pm 0.03    & 34711/20 & 2024.91 \pm 0.03    & 1074/22 & 2092.26 \pm 0.06                 & 229887/43 & 751091 /43 &  3.89\times10^{-15}\\
\rowcolor{lightgray} \rownumber & NGC~1068 & \nodata & \nodata             & \nodata     & \nodata          & \nodata & \nodata                          & \nodata   & \nodata    &  \nodata           \\
                     \rownumber & NGC~1320 & 0.57 \times 0.50 &    1.30 \pm 0.01    &    60/24 &    1.02 \pm 0.01    &   31/20 &    1.09 \pm 0.01                 &     85/45 & 4/45       &  1.59\times10^{-01}\\
\rowcolor{lightgray} \rownumber & NGC~2110 & 0.63 \times 0.52 &   63.32 \pm 0.01    &   648/22 &   71.41 \pm 0.03    &   13/14 &   67.45 \pm 0.02                 &    638/37 & 49/37      &  1.01\times10^{-01}\\
\rowcolor{lightgray} \rownumber & NGC~2782 & 0.62 \times 0.47 &    2.54 \pm 0.01    &    41/28 &    1.67 \pm 0.01    &   38/30 &    2.08 \pm 0.01                 &     86/59 & 28/59      &  4.19\times10^{-05}\\
                     \rownumber &  IC~2461 & 0.48 \times 0.44 &    0.53 \pm 0.01    &    36/26 &    0.46 \pm 0.01    &   43/30 &    0.45 \pm 0.01                 &     72/57 & 1/57       &  4.62\times10^{-01}\\
\rowcolor{lightgray} \rownumber & NGC~2992 & 0.86 \times 0.48 &    9.05 \pm 0.01    &   111/25 &    7.70 \pm 0.02    &  127/20 &    7.84 \pm 0.02                 &    216/46 & 176/46     &  1.92\times10^{-07}\\
                     \rownumber & NGC~3081 & 0.92 \times 0.45 &    1.06 \pm 0.01    &    27/26 &    0.84 \pm 0.01    &   29/27 &    0.95 \pm 0.01                 &     57/54 & 9/54       &  5.04\times10^{-03}\\
                     \rownumber & NGC~3089 & 0.81 \times 0.40 & \nodata             & \nodata  & \nodata             & \nodata &    0.23 \pm (<0.01)^*            &      0/2  & \nodata    &  \nodata           \\
\rowcolor{lightgray} \rownumber & NGC~3079 & 0.72 \times 0.48 &  171.07 \pm 0.01    &    56/26 &  161.32 \pm 0.01    & 1535/30 &  168.64 \pm 0.02                 &  21462/57 & 200477/57  &  1.11\times10^{-16}\\
                     \rownumber & NGC~3227 & 0.58 \times 0.45 &   11.23 \pm 0.01    &   308/28 &    6.87 \pm 0.01    &   60/30 &    8.57 \pm 0.01                 &    343/59 & 47/59      &  6.25\times10^{-03}\\
                     \rownumber & NGC~3786 & 0.36 \times 0.32 & \nodata             & \nodata  &    0.93 \pm (<0.01) &   92/29 &    0.93 \pm 0.01^{\dagger}       &     91/29 & 1/29       &  6.26\times10^{-01}\\
\rowcolor{lightgray} \rownumber & NGC~4151 & 0.48 \times 0.46 &   58.39 \pm 0.01    &   774/26 &   45.45 \pm 0.01    &  341/30 &   50.59 \pm 0.01                 &    850/57 & 302/57     &  3.44\times10^{-05}\\
                     \rownumber & NGC~4180 & 0.52 \times 0.51 &    0.61 \pm 0.01    &    32/25 & \nodata             & \nodata &    0.61 \pm 0.01^{\ddagger}      &     32/25 & 0/25       &  7.71\times10^{-01}\\
\rowcolor{lightgray} \rownumber & NGC~4235 & 0.51 \times 0.48 &    6.24 \pm 0.01    &   153/25 &    8.70 \pm (<0.01) &  308/28 &    7.65 \pm 0.01                 &    582/54 & 3819/54    &  1.11\times10^{-16}\\
\rowcolor{lightgray} \rownumber & NGC~4388 & 0.58 \times 0.49 &    3.59 \pm 0.02    &    32/25 &    2.99 \pm 0.01    &   73/30 &    3.32 \pm 0.02                 &    109/56 & 7/56       &  6.63\times10^{-02}\\
\rowcolor{lightgray} \rownumber & NGC~4593 & 0.42 \times 0.38 &    1.43 \pm 0.02    &    15/12 &    1.34 \pm 0.01    &   88/28 &    1.40 \pm 0.01                 &     95/41 & 9/41       &  5.20\times10^{-02}\\
\rowcolor{lightgray} \rownumber & NGC~5290 & 0.48 \times 0.43 &    6.93 \pm 0.01    &    43/26 &    6.68 \pm 0.01    &   54/30 &    6.79 \pm 0.02                 &     80/57 & 184/57     &  2.22\times10^{-16}\\
\rowcolor{lightgray} \rownumber & NGC~5506 & 0.86 \times 0.46 &  118.81 \pm 0.01    &  4333/20 &   81.64 \pm 0.01    & 2244/20 &   95.68 \pm 0.02                 &   7182/41 & 7883/41    &  4.26\times10^{-08}\\
                     \rownumber & NGC~5899 & 0.52 \times 0.45 &    2.83 \pm 0.01    &    29/26 &    1.85 \pm 0.01    &   63/30 &    2.19 \pm 0.01                 &     91/57 & 7/57       &  4.12\times10^{-02}\\
                     \rownumber & NGC~6814 & 0.92 \times 0.47 &    1.13 \pm 0.01    &    91/21 &    0.64 \pm 0.01    &   44/14 &    0.96 \pm 0.01                 &    166/36 & 5/36       &  2.97\times10^{-01}\\
                     \rownumber & NGC~7314 & 0.89 \times 0.42 &    1.72 \pm (<0.01) &    27/27 &    0.95 \pm 0.01    &   36/30 &    1.37 \pm 0.01                 &     86/58 & 164/58     &  4.00\times10^{-15}\\
                     \rownumber & NGC~7378 & 0.50 \times 0.36 & \nodata             & \nodata  & \nodata             & \nodata &    0.10 \pm (<0.01)^*            &      0/2  & \nodata    &  \nodata           \\
                     \rownumber & NGC~7465 & 0.57 \times 0.45 &    1.14 \pm 0.01    &   145/28 &    0.77 \pm 0.01    &   42/30 &    0.89 \pm 0.01                 &    165/59 & 22/     59 &  7.23\times10^{-03}\\
                     \rownumber & NGC~7479 & 0.46 \times 0.45 &    2.46 \pm 0.01    &    66/26 &    2.27 \pm 0.01    &   66/30 &    2.21 \pm 0.01                 &    115/57 & 5/      57 &  1.26\times10^{-01}
\enddata
\tablecomments{Fitting results for peak flux densities from channelized data for the VLA FRAMEx sample. Highlighted rows indicate VLBA-detected sources. \textbf{Column 1.} target name. \textbf{Column 2.} FWHM of restoring beam from matched resolution image cube. \textbf{Column 3.} fitted peak flux density at 6 GHz from $4-8$ GHz power-law fit. \textbf{Column 4.} $\chi^2$ from $4-8$ GHz power-law fit over the degrees of freedom. \textbf{Column 5.} fitted peak flux density at 10 GHz from $8-12$ GHz power-law fit. \textbf{Column 6.} $\chi^2$ from $8-12$ GHz power-law fit over the degrees of freedom. \textbf{Column 7.} fitted peak flux density at 8 GHz from $4-12$ GHz power-law with curvature fit; the $8-12$ GHz data were corrected by the factor noted in Column 6 of Table \ref{tab:spix}. \textbf{Column 8.} $\chi^2$ from $4-12$ GHz power-law with curvature fit over the degrees of freedom \textbf{Column 9}. difference of $\chi^2$ values from the power law versus power law with curvature fitting results over the degrees of freedom used in the F-test. \textbf{Column 10}. $p$-value of the F-test. Null hypothesis was rejected in the case of a 99.73\% confidence interval.\\
$^{*}$NGC~3089 and NGC~7378 were not fit with curvature.\\
$^{\dagger}$NGC~3786 was fit with a reference frequency set to 10 GHz, so the fitted peak flux density is measured at 10 GHz.\\
$^{\ddagger}$NGC~4180 was fit with a reference frequency set to 6 GHz, so the fitted peak flux density is measured at 6 GHz.}
\label{tab:chi2fits}
\end{deluxetable}
\end{rotatetable*}

\end{document}